*Chatain et al., version accepted in Icarus (2020)*

# Interaction dust – plasma in Titan's ionosphere: an experimental simulation of aerosols erosion

A. Chatain (1,2,*), N. Carrasco (1), N. Ruscassier (3), T. Gautier (1), L. Vettier (1) and O. Guaitella (2)

(1) LATMOS/IPSL, UVSQ, Université Paris-Saclay, Sorbonne Université, CNRS, 78280 Guyancourt, France
(2) LPP, École polytechnique, Sorbonne Université, université Paris-sud, CNRS, 91128 Palaiseau, France
(3) LGPM, Centrale-Supélec, 91190 Gif-sur-Yvette, France

* audrey.chatain@latmos.ipsl.fr

ABSTRACT

Organic aerosols accumulated in Titan's orange haze start forming in its ionosphere. This upper part of the atmosphere is highly reactive and complex ion chemistry takes place at altitudes from 1200 to 900 km. The ionosphere is a nitrogen plasma with a few percent of methane and hydrogen. Carbon from methane enables the formation of macromolecules with long organic chains, finally leading to the organic aerosols. On the other hand, we suspect that hydrogen and the protonated ions have a different erosive effect on the aerosols.

Here we experimentally studied the effect of hydrogen and protonated species on organic aerosols. Analogues of Titan's aerosols were formed in a radiofrequency capacitively coupled plasma (RF CCP) discharge in 95% $N_2$ and 5% $CH_4$. Thereafter, the aerosols were exposed to a DC plasma in 99% $N_2$ and 1% $H_2$. Samples were analysed by scanning electron microscopy and *in situ* infrared transmission spectroscopy. Two pellet techniques – KBr pressed pellets and thin metallic grids - were compared to confirm that modifications seen are not due to the material used to make the pellet.

We observed that the spherical aerosols of ~500 nm in diameter were eroded under $N_2$-$H_2$ plasma exposure, with the formation of holes of ~10 nm at their surface. Aerosols were globally removed from the pellet by the plasma. IR spectra showed a faster disappearance of isonitriles and/or carbo-diimides compared to the global band of nitriles. The opposite effect was seen with β-unsaturated nitriles and/or cyanamides. Double bonds as C=C and C=N were more affected than amines and C-H bonds. N-H and C-H absorption bands kept a similar ratio in intensity and their shape did not vary.

Therefore, it seems that carbon and hydrogen play opposite roles in Titan's ionosphere: the carbon from methane leads to organic growth while hydrogen and protonated species erode the aerosols and react preferentially with unsaturated chemical functions.

# 1. Introduction

Atmospheric aerosols can have important impacts on a planet climate. They are efficient scatterers of light and nucleation sites for clouds. Consequently, they can modify strongly the atmospheric radiative budget and impact dynamics in a way hard to predict. These particles also play a major role in atmospheric chemistry, especially as they favour surface-catalysed processes. While it is difficult to understand and quantify their effects, atmospheric aerosols are found in all planetary atmospheres, and this strengthens the necessity to study them. In the various planetary atmospheres of the solar system, aerosols come from very different origins, such as volcanism, oceans or human actions on Earth (Jacobson et al., 2000; Pöschl, 2005), dust on Mars (Madeleine et al., 2011) or atmospheric chemistry on Jupiter and Saturn (Courtin, 2005; Guerlet et al., 2015; Zhang et al., 2015).

Titan is an extreme example in the Solar system where aerosols are present in huge quantity in the atmosphere and hide the surface from space in many wavelengths. The understanding of the moon's atmosphere is therefore strongly limited by our knowledge of the particle interactions with the surrounding species and radiations. The Cassini-Huygens mission discovered that Titan's aerosols start forming in the upper atmosphere, above 1200 km (Waite et al., 2007). Titan's ionosphere is an environment particularly adapted for this purpose. Indeed, nitrogen and methane molecules present in low density are submitted to solar radiation and energetic particles from Saturn's magnetosphere. Excited and charged species are formed and start chemical reactions leading to the formation of heavy negative ions (Coates et al., 2007; Westlake et al., 2014), and later on to dust particles (Lavvas et al., 2013). The ionosphere is consequently a dusty plasma where organic aerosols are mixed among plasma species.

Studies in plasma research show that organic material is usually modified when exposed to an ionized environment (D'Agostino et al., 1997). Depending on the plasma characteristics, the surface can be etched or covered by a deposit, e.g. in microelectronics (Mutsukura and Akita, 1999). Many applications also use plasmas to modify physical and chemical properties of surfaces, such as reactivity, wettability (Šíra et al., 2005) or bio-compatibility (Truica-Marasescu and Wertheimer, 2008). Therefore, it is likely that Titan organic aerosols constantly interact with plasma species during their life in the ionosphere, leading to growth, erosion and chemical modifications.

Laboratory studies have already been performed to mimic a possible evolution of organic matter in protoplanetary discs and interstellar medium. Especially, the effects of ion, electron and photon irradiation (Kuga et al., 2015; Laurent et al., 2014; Öberg, 2016) are processes similar to what could happen in Titan's ionosphere. A first work was done to study the evolution of analogues of aerosols (called "tholins") exposed to VUV radiations (Carrasco et al., 2018). Under exposure, the nitrile infrared bands of tholins shift and their intensities are modified. The impact of atomic hydrogen on tholins has also been studied in Sekine et al., 2008. No modification of the chemical structure of tholins other than hydrogenation have been observed.

The next step to understand the evolution of aerosols in Titan's ionosphere is to focus on the effect of plasma species, and especially electrons and ions. Aerosols stay several days in the ionospheric layer ionized by VUV radiations and magnetospheric electrons from 1200 to 900 km (Lavvas et al., 2013). The effect of plasma exposure on such time scale on Titan is still unknown. In this work, we investigate for the first time the effect of plasma exposure on Titan aerosols analogues. Both morphology and chemical properties are investigated by the mean of electron microscopy and *in situ* infrared spectroscopy.

# 2. Experimental setup and protocol

## 2.1 Sample synthesis with PAMPRE reactor

The first step of this experimental campaign was to produce analogues of Titan aerosols. The PAMPRE experiment located in LATMOS is especially suited to provide such samples (Szopa et al., 2006). Its specificity is to create analogues in suspension in the chamber using a radiofrequency (RF) capacitively coupled plasma (CCP) discharge. The organic aerosols are formed in the middle of the plasma and sediment once they reach a limit size. Their retention time in the plasma can be modified by parameters of the experiment (Hadamcik et al., 2009).

The chamber is a stainless steel cylinder of 40 cm in height and 30 cm in diameter. The upper electrode is a stainless steel disk grid of 12.6 cm in diameter. An aluminium alloy cylindrical box is adjusted to the upper electrode to confine the plasma. Both the upper electrode and the bottom of the box are pierced with small holes to let the gas go through. A RF potential is applied to the upper electrode, leading to the lightning of the plasma discharge in the box.

As this work simply intend to test the reactivity of tholins exposed to plasma conditions without any further constraints, we used usual working conditions in PAMPRE to form the analogues (Sciamma-O'Brien et al., 2010). They reproduce the average methane concentration and lead to an electron energy distribution similar to the solar spectrum in Titan's ionosphere (Szopa et al., 2006). Therefore, 0.90 mbar of a $N_2$-$CH_4$ mixture at 5% in methane was injected at 55 sccm (standard cubic centimetres per minute). We used a pre-mixed gas bottle with 10% methane in nitrogen (Air Liquide – CRYSTAL mixture), and diluted it with pure nitrogen (Air Liquide – alphagaz 2). We used high purity gases (> 99.999%). The discharge was ignited during several hours with 30 W incident power. The temperature of the polarized electrode reaches 60-80°C during the production.

In these conditions, particles stay one to two minutes in the plasma and grow up to a few hundred nanometres (Hadamcik et al., 2009). After the production, tholins are temporarily exposed to ambient air during their collection and the making of pellets. Water adsorption and oxidation should appear on the first nanometres of the grains, as measured in (Carrasco et al., 2016). Before each experiment, samples are exposed to vacuum and heated to desorb water (detailed in section 2.4).

## 2.2 Exposure in a DC plasma reactor

In order to approach ionospheric environmental conditions in the laboratory, a great care was given to try to compensate effects to be as representative to Titan as possible (see Table 1). Especially, in the plasma discharge the laboratory pressure cannot be as low as on Titan's ionosphere ($10^{-8}$ to $10^{-6}$ mbar). But if the electron density is also increased in the experiment, the ionization degree can be the same as on Titan. Nevertheless, as the electron density is $10^7$ times higher in the experiment, erosion processes should be ~$10^7$ times faster (days become seconds). Therefore, the exposure time is reduced in the laboratory compared to Titan's case. Finally, Titan ionospheric aerosols are smaller than tholins formed in PAMPRE. As the erosion of bigger grains takes more time, the exposure time should be adapted to compensate this difference. We therefore expose grains during minutes to hours in the laboratory.

| | on Titan | in the experiment |
|---|---|---|
| pressure (mbar) | $10^{-8}$ to $10^{-6}$ | 4 |
| electron density $n_e$ ($cm^{-3}$) | [700 ; 3000] varying with the altitude (Ågren et al., 2009) | ~$10^{10}$ |
| ionization degree | [$10^{-8}$ ; $10^{-6}$] varying with the altitude | ~$10^{-7}$ |
| $\alpha = \frac{n_e}{n_{tot}}$ | | |
| grain size (nm) | 1 to a few 10s (Lavvas et al., 2013) | a few 100s (Hadamcik et al., 2009) |
| exposure time | a few days (Lavvas et al., 2013) | a few minutes to hours |
| hydrogen amount | ~0.5% $H_2$ and 1.4% $CH_4$ | 1% $H_2$ |

*Table 1: Comparison of physical variables fundamental to the study of aerosol evolution by plasma in the ionosphere of Titan and in the experiment described here.*

We exposed the samples in the positive column of a DC glow discharge. It is a kind of discharge often used for the analysis of plasma-surface interactions (Azzolina-Jury and Thibault-Starzyk, 2017). The main reason is its high homogeneity which enables to attribute sample modifications to precise characteristics of the plasma. The plasma was ignited in a 23 cm long and 2 cm inner diameter pyrex tube in which the sample was deposited on a glass support. The current was set at 20 mA, which gives an electron density in the tube of about $10^{10}$ $cm^{-3}$ and a gas temperature that does not exceed 330K.

Concerning the gas mixture, the objective was to mimic the composition of Titan's ionosphere but also to avoid the formation of new particles during the exposure, to differentiate the processes of formation from those of evolution of aerosols in the ionosphere. Therefore, to prevent organic growth, methane was not introduced in the mixture and only nitrogen and hydrogen were used. In Titan's ionosphere, $H_2$ is present in proportions of ~0.4% around 1000 km of altitude (Cui et al., 2008). Besides, hydrogen atoms used for chemical processes can also come from $CH_4$ (~1.4%) which is easily dissociated in the ionosphere of Titan (P. Lavvas et al., 2011; Sciamma-O'Brien et al., 2010; Vuitton et al., 2019). The exact percentage of hydrogen active for physical and chemical evolution in the plasma is not fully constrained to date. We chose a gas mixture of 99% $N_2$ and 1% $H_2$ for our plasma discharge. We used high purity gases (> 99.999%). The gas from a 95% $N_2$ – 5% $H_2$ gas bottle (Air Liquide – CRYSTAL mixture) was diluted in pure $N_2$. We injected 5 sccm of gas and adjusted the pressure to 4 mbar.

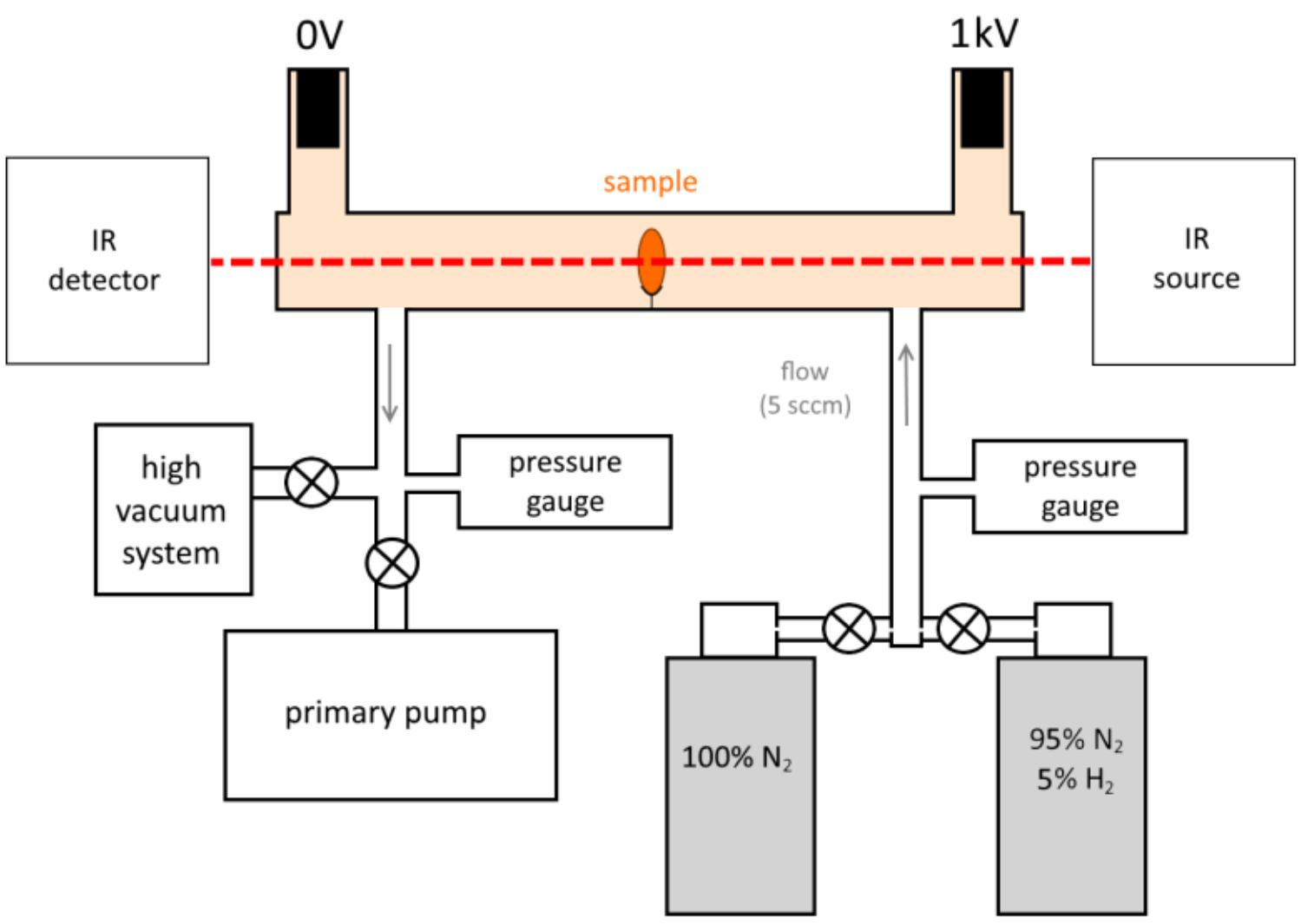

*Figure 1: Experimental setup: exposure of tholin samples in a glow discharge monitored by in situ IR spectroscopy*

## 2.3 Pellet analysis with SEM and IR spectroscopy

### 2.3.1 Surface evolution by SEM

The surfaces of different samples were compared before and after plasma exposure by means of Scanning Electron Microscopy (SEM). The instruments used were an environment SEM Quanta 200 from FEI (with a LFD detector, a high voltage of 12.5kV and at 0.5 mbar) and a Field Emission Gun SEM LEO1530 (with a high voltage of 2kV). Scanning electron microscopy gives information on the topography of sample surfaces. An EDS (Energy Dispersive X-ray Spectroscopy) probe coupled to the environment SEM also determined the elementary composition at different locations on the surface.

### 2.3.2 Absorption bands modifications by IR spectroscopy

Infrared transmission spectroscopy was used to get information on the chemical evolution of the pellets, with a similar technique as in (Jia and Rousseau, 2016) and in (Mohammad Gholipour et al., 2017). The reactor was put inside the sample compartment of a Nicolet 6700 FTIR from Thermo so that the evolution of the transmitted light could be monitored before, during and after the exposure. The spectra were acquired by a DTGS detector between 900 and 4000 $cm^{-1}$, this range being limited by the KBr and $CaF_2$ windows used respectively on the FTIR and the reactor. We chose a spectral resolution of 1 $cm^{-1}$, a Michelson speed of 0.63 cm/s and averaged each spectrum from 5 scans. With these parameters, spectra were acquired in ~30s. We aimed such a short acquisition time to have also a short time step resolution during the evolution of the sample. The width of the IR beam at the focal point is a few millimetres.

## 2.4 Realisation of two different kinds of pellets

Tholins formed in PAMPRE are grains of few hundred nanometres, likely to be dragged by the gas flow and/or the electrostatic force when exposed to a DC plasma. Here we compare two different techniques to keep them in the plasma during the exposure (see Table 2). A major constraint is that the final sample should not be opaque in IR wavelengths.

The first technique is the most common (Abdu et al., 2018). We made pressed pellets with tholins and a KBr matrix. A pressure of 5 tons was applied to form pellets of 0.13 g, 1.3 cm in diameter and about 0.5 mm in thickness. A mass mixing ratio of 1.5% tholins - 98.5% KBr thin grains has been found optimal for both mechanical solidity and IR transmission. KBr is totally transparent in IR wavelengths. However, the KBr grains of the pellet are suspected to protect tholins from the plasma species. Besides, KBr is known to be strongly hydrophilic. Samples have to be heated under vacuum before the beginning of the experiments to outgas adsorbed water.

To be certain that KBr grains do not affect our results, we compared them to those obtained with another technique. 1 mg of tholin grains are spread on a thin stainless steel grid of 1.3 cm in diameter. Threads are 25 µm-large and meshes are of 38 µm. This leads to a transmission of 20% in IR wavelengths [cf Figure 11 in S.I.].

Before each experiment (for both KBr and grid pellets) samples are put inside the glass reactor under secondary pumping for a few hours, reaching a pressure of ~$2.10^{-6}$ mbar. At the beginning of the pumping, the glass, and the sample in contact with the glass, are heated for 45 min at 80 to 100°C to desorb water. The effect of temperature on tholins is discussed in section 3.4.2. The IR spectra of tholins during heating show the disappearance of adsorbed water, but no evolution is seen on the other absorption bands. Spectra are given in S.I. (see section 6.6) and are discussed with the analysis of each band (see sections 3.3.2-3-4).

| Pellet | Characteristics |
|---|---|
| $P_{th}$ | KBr pellet with 1.5% tholins (1.6 mg of tholins), 110 mg, pressed under 5 tons |
| $P_{kbr}$ | pure KBr, 115 mg, pressed under 5 tons |
| $P_{gr}$ | 1 mg of tholins spread on a thin metallic grid. Experiment done twice for reproducibility. |

*Table 2: Studied samples*

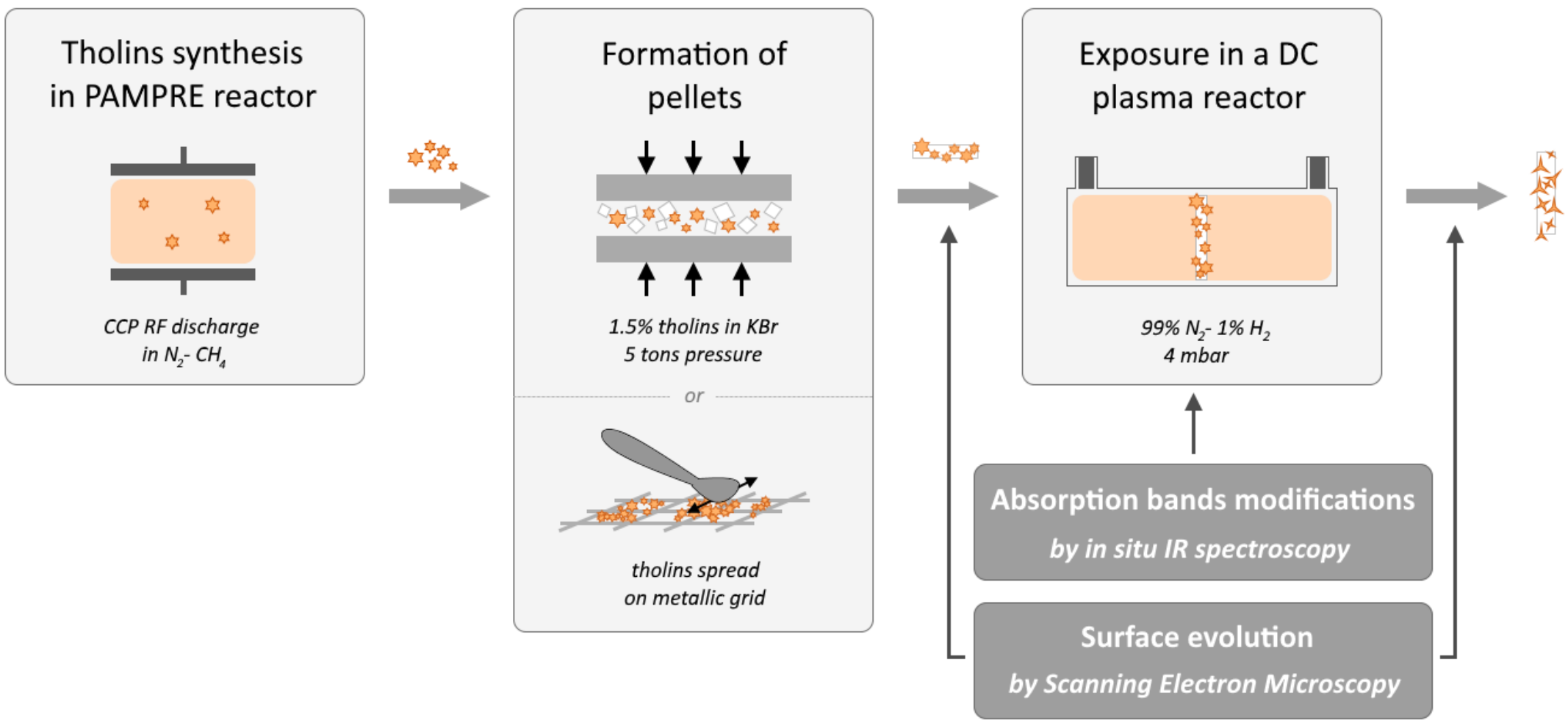

*Figure 2: Experimental protocol*

# 3. Results

## 3.1 Characterization of the samples before exposure

### 3.1.1 IR absorption spectrum of tholins

Two different forms of tholins grow in the PAMPRE reactor: powders in suspension in the gas phase and films deposited on surfaces. Mainly tholins films have been previously studied by IR transmission spectroscopy (Gautier et al., 2012). Even if formed at the same time in the reactor PAMPRE, powders and films follow different production pathways. This leads to different elementary composition. In particular, powders have a higher nitrogen content than films (Carrasco et al., 2016).

The IR signature of tholins shows three major absorption bands: from 2700 to 3600 $cm^{-1}$ due to amine and C-H stretching bonds, around 2200 $cm^{-1}$ due to nitriles and isonitriles stretching modes and from 1100 to 1800 $cm^{-1}$ due to many different stretching and bending bonds (Imanaka et al., 2004).

The comparison of powders and film spectra (see Figure 3) shows a few differences: nitriles (C≡N) and C-H bonds (~2950 $cm^{-1}$) have lower contributions in powders than in films, which is consistent with the work cited above (Carrasco et al., 2016). One study has already compared IR spectra of powders and films (Quirico et al., 2008). It was done with a different percentage of methane than in the present work but the observations were similar.

Normalized IR spectra of the same tholin sample, but obtained by the two different methods on powders (with KBr *vs* on grids) are only slightly different (see Figure 3). The powders spectra in Figure 3 will serve as references for the following study of powders evolution under plasma exposure.

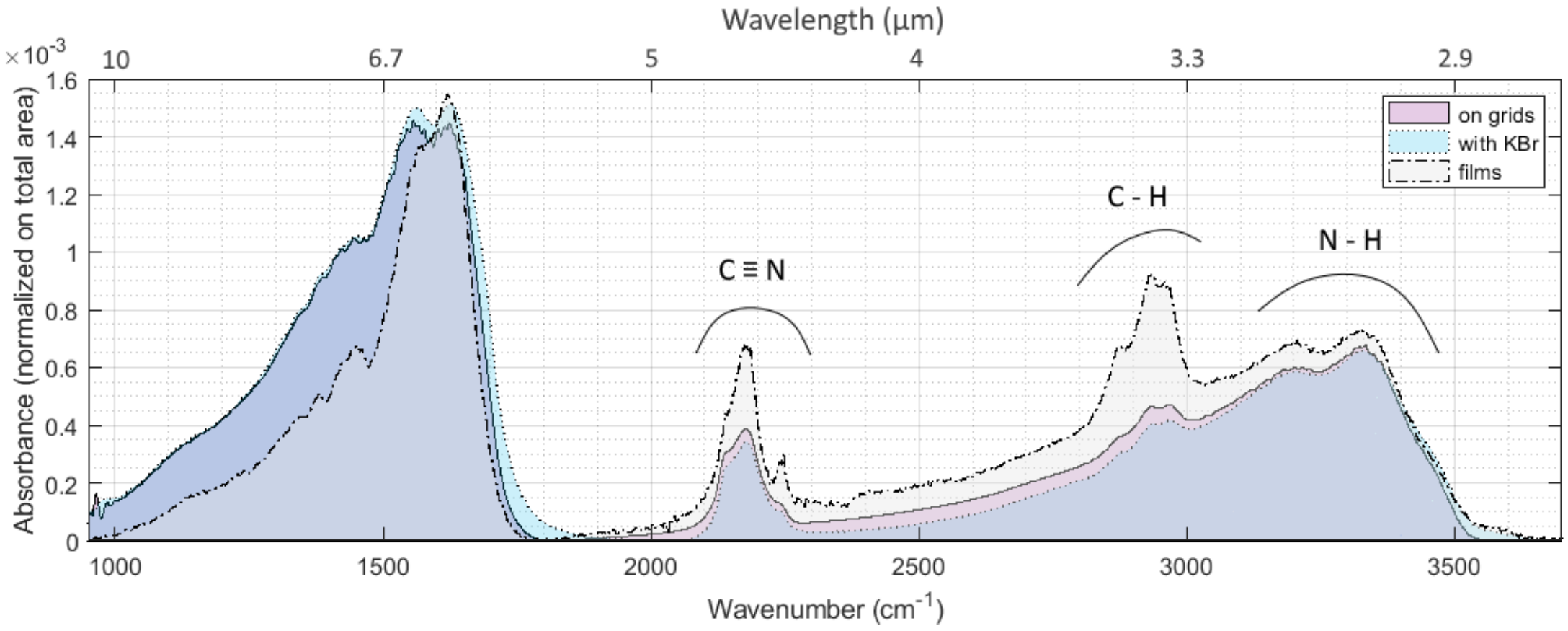

*Figure 3: IR spectra of tholins formed in PAMPRE in 95% $N_2$ - 5% $CH_4$. Absorbance is normalized on the total area of the spectra from 950 to 3700 $cm^{-1}$. Superimposition of the IR signature of powders and films from* (Gautier et al., 2012) *(dotted-dashed line). Spectra with the two techniques presented here: pellets with grids (plain line) and pellets with KBr (dotted line).*

### *3.1.2 SEM pictures of the KBr pellet surface (method 1)*

SEM pictures of the surfaces of pellets were taken for tholins-KBr and pure KBr pellets at different scales (from 500 µm to 10 µm – cf Table 4). All the surfaces of samples were analysed and the pictures presented here are representative of the whole samples.

An EDS probe was used to obtain composition maps of the surface (cf Figure 4). Pictures clearly show two different materials. White grains are made of potassium and bromine as expected, and the black material contains carbon, nitrogen and some oxygen. Oxygen is not present in the gas injected to form tholins in PAMPRE, but previous studies showed that they oxidize as soon as they are in contact with air (Carrasco et al., 2016). KBr is highly hydrophilic and its mixture with tholins could also enhance the oxidation of tholins.

The mixing of tholins and KBr is not homogeneous at the micrometric scale. KBr grains are ~200 µm large. Tholins are small spheres of ~500 nm in diameter, but they tend to aggregate in clusters of 10 to 100 µm at the surface of KBr grains or in between grains. KBr has been flattened by the tableting. Individual tholins grains are still round, but clusters are spread out and their flat surfaces are cracked.

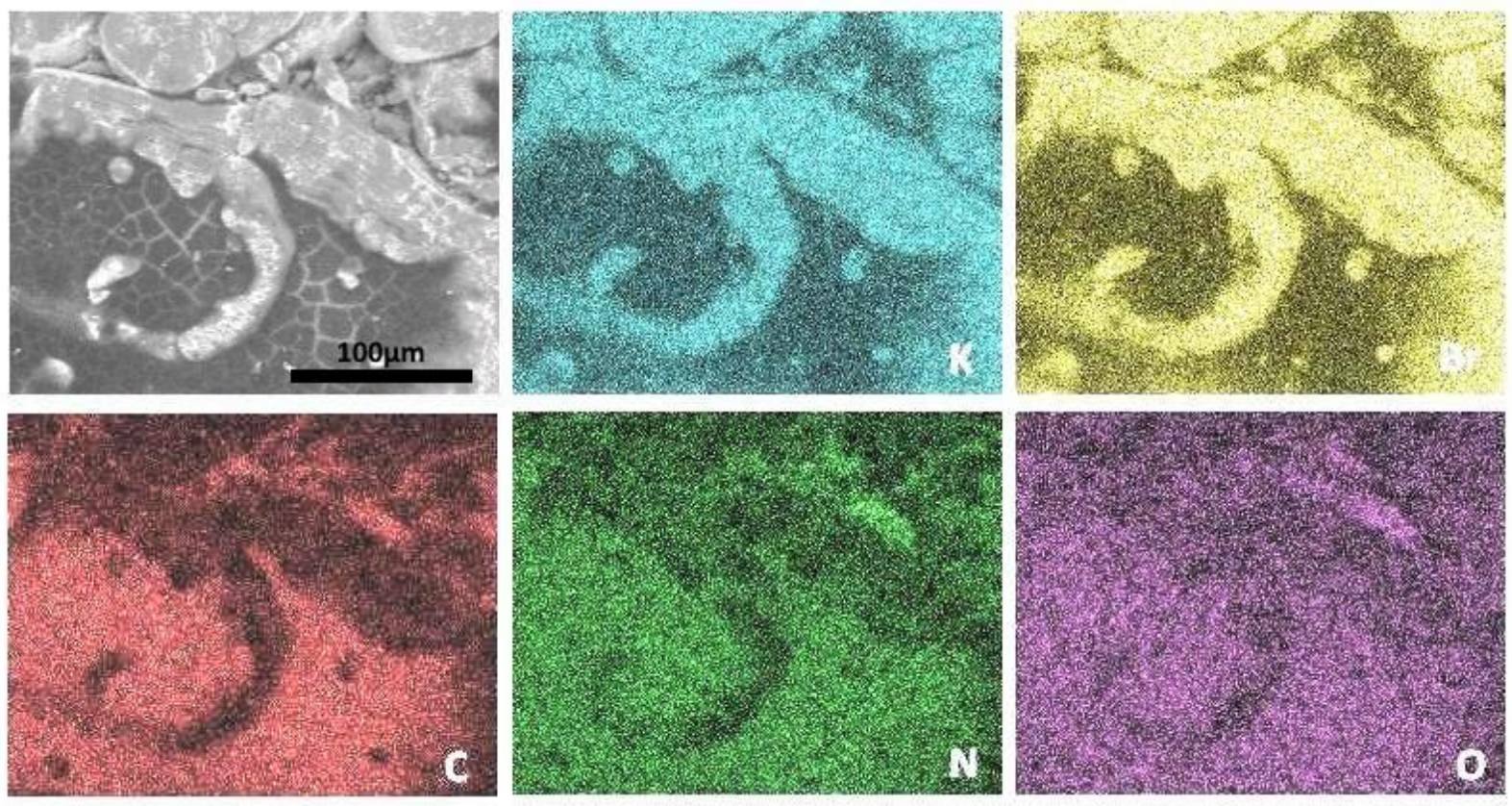

*Figure 4: Elementary analysis with an EDS probe of the surface of a pellet with KBr and tholins ($P_{th}$). Potassium, bromine, carbon, nitrogen and oxygen have been detected. Hydrogen cannot be seen by this method. The 100 µm scale is similar for all six images.*

### 3.1.3 SEM pictures of the tholins spread on grids (method 2)

Tholins gather inside the meshes. They form round grains of 450-500 nm in diameter with a rather smooth surface, similarly as seen in Hadamcik et al., 2009. SEM pictures are given in Table 5.

## 3.2 Physical erosion of the pellets

### 3.2.1 Evolution of the surfaces

Naked eyes are enough to see that the sample changes during exposure (see Table 3). Initial brown samples in KBr ($P_{th}$) turn into white, become rougher and eroded on the sides. Tholins are also removed from the grids ($P_{gr}$). ~1 mg of tholins is used in $P_{th}$ and $P_{gr}$. Tholins are still present in $P_{th}$ after 4h plasma while all tholins are removed from $P_{gr}$ in less than 1h. KBr grains protect tholins from erosion.

| | 1.5% tholins in KBr ($P_{th}$) | pure KBr ($P_{kbr}$) | tholins spread on grid ($P_{gr}$) |
|---|---|---|---|
| Before exposure | 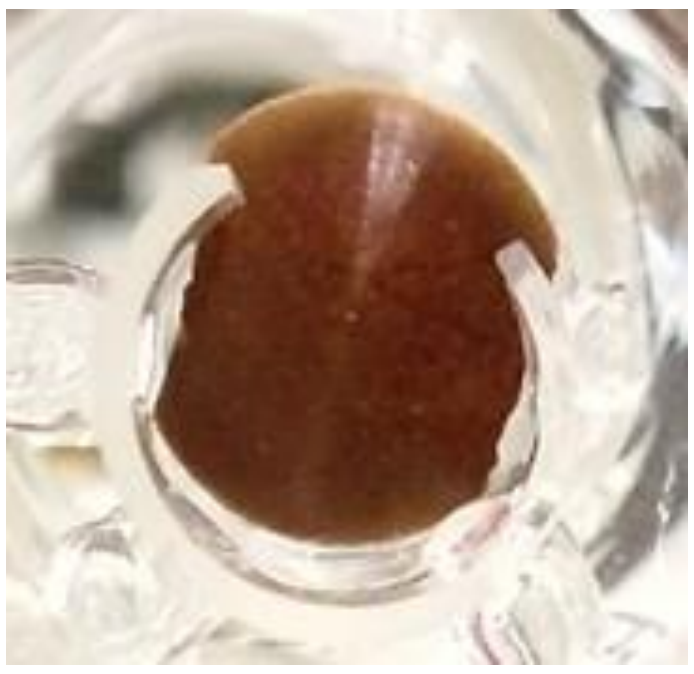 | 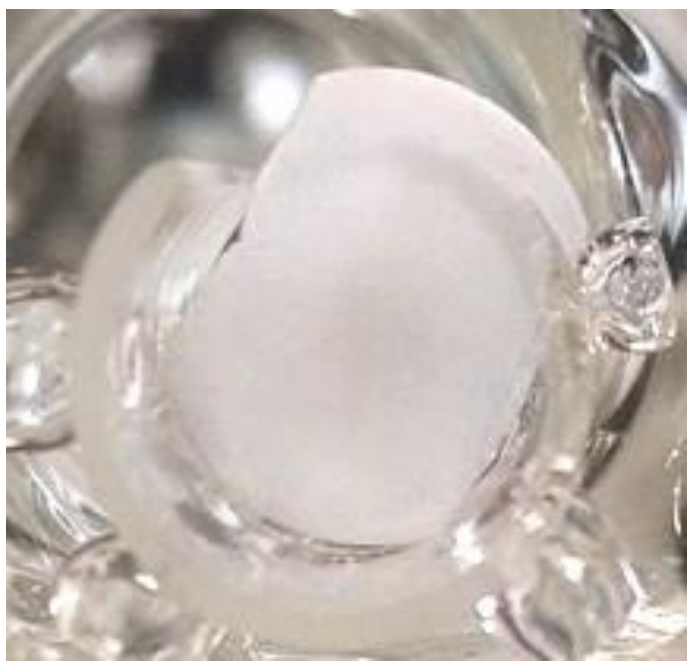 | 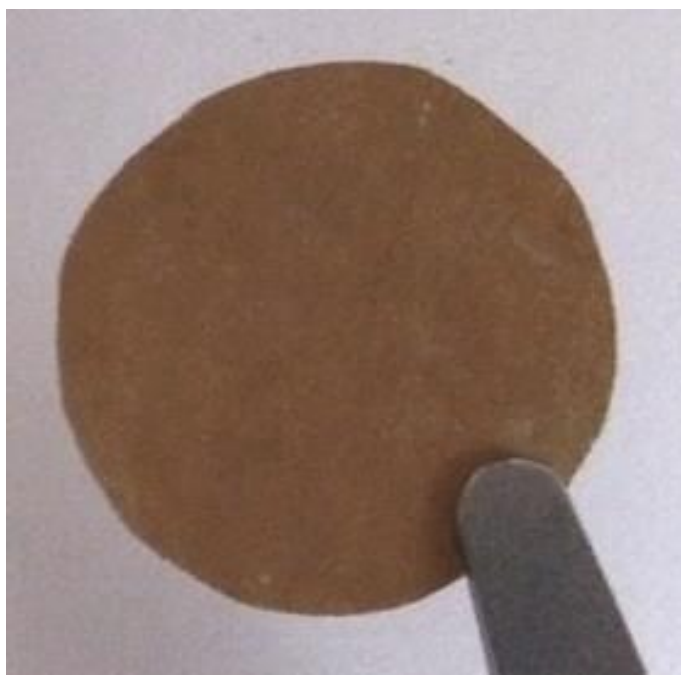 |
| After exposure | 5h 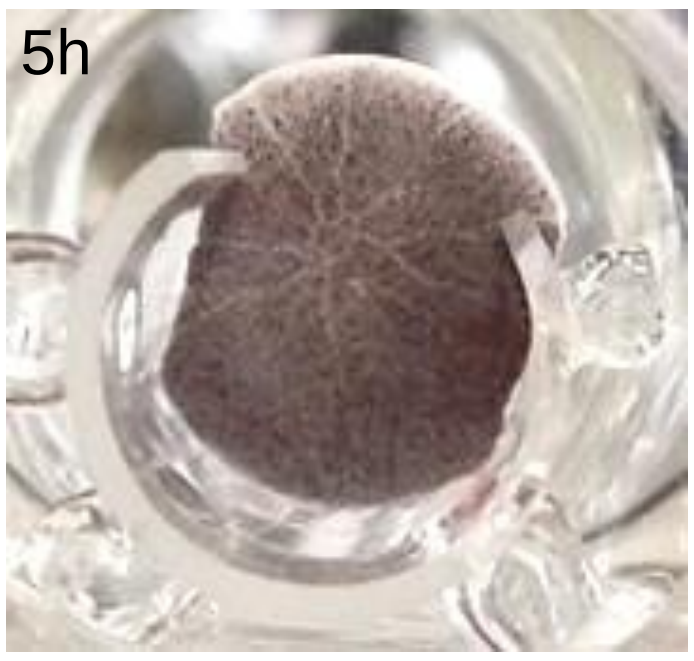 | 3h 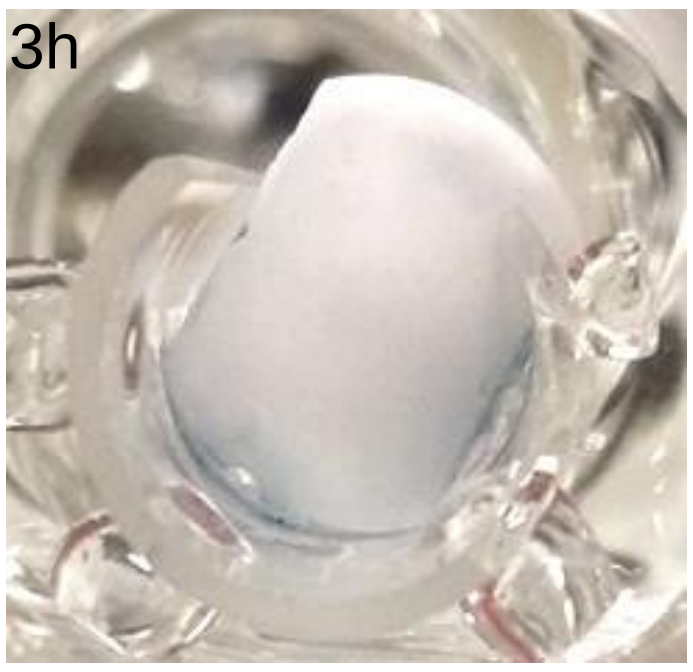 | 1h 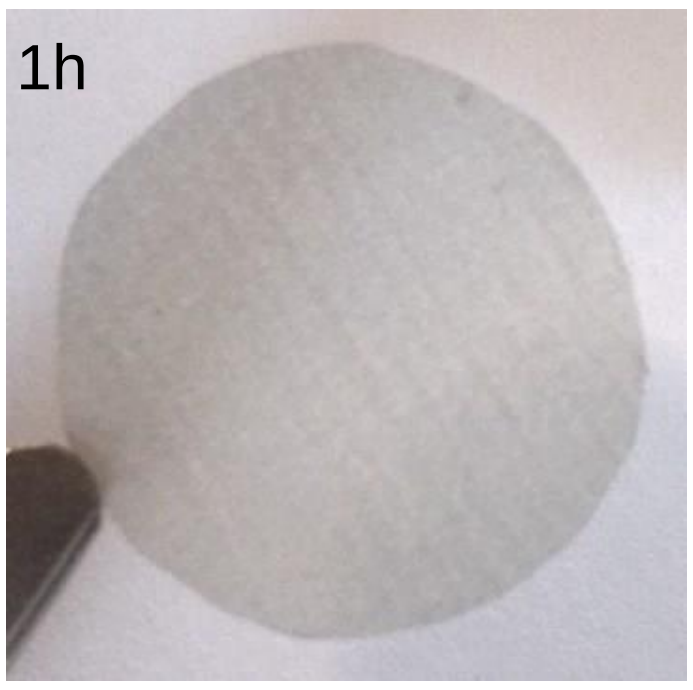 |

*Table 3: Pictures of the samples inside their holder before and after being exposed to plasma. Samples with 1.5% of tholins in KBr ($P_{th}$), with pure KBr ($P_{kbr}$) and tholins spread on a metallic grid ($P_{gr}$).*

Concerning the KBr(-tholins) pellets, SEM pictures of the pellets surfaces after exposure show a strong physical erosion (see Table 4). Sputtering indentations appear on KBr grains. Tholins are totally removed from the surface after a few hours of exposure to the plasma. However, while only KBr can be seen at the surface, the pellets appear still brown. Tholins are still present inside, but KBr grains protect them from a fast full erosion. Therefore, the use of KBr to form pellets slows down the erosion of tholins by forming a protecting layer at the surface of the pellet.

| | $P_{th}$ | $P_{kbr}$ |
|---|---|---|
| **before exposure** | 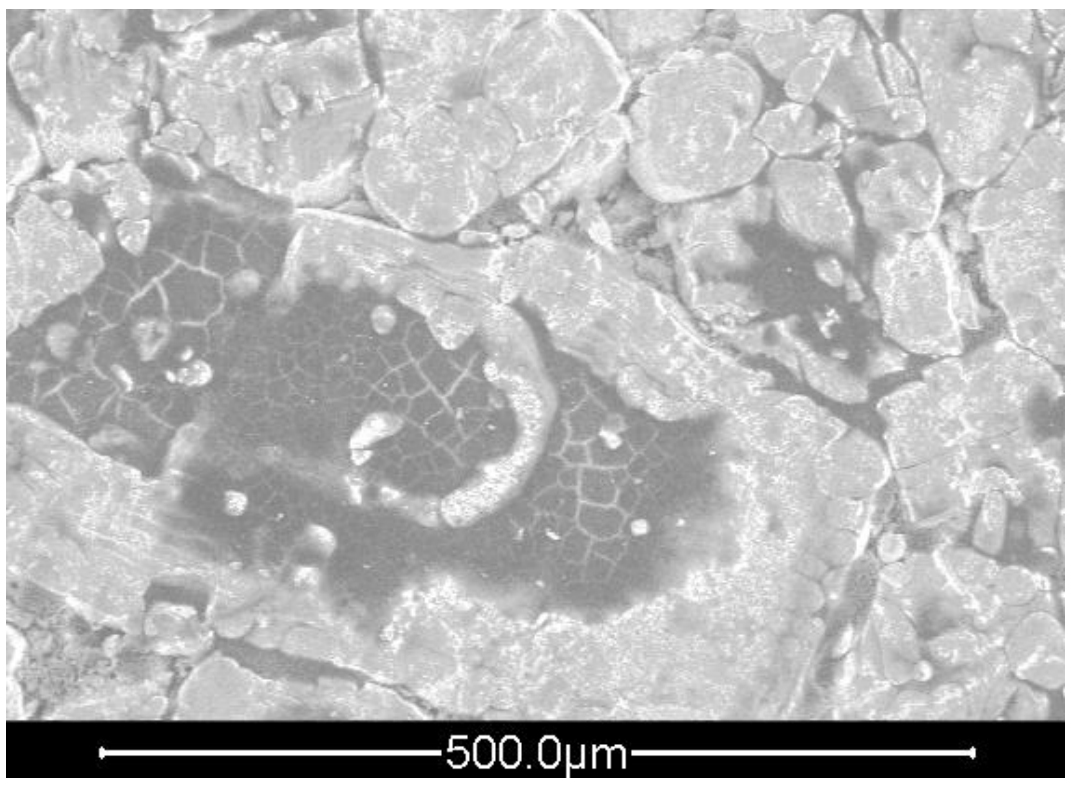  | 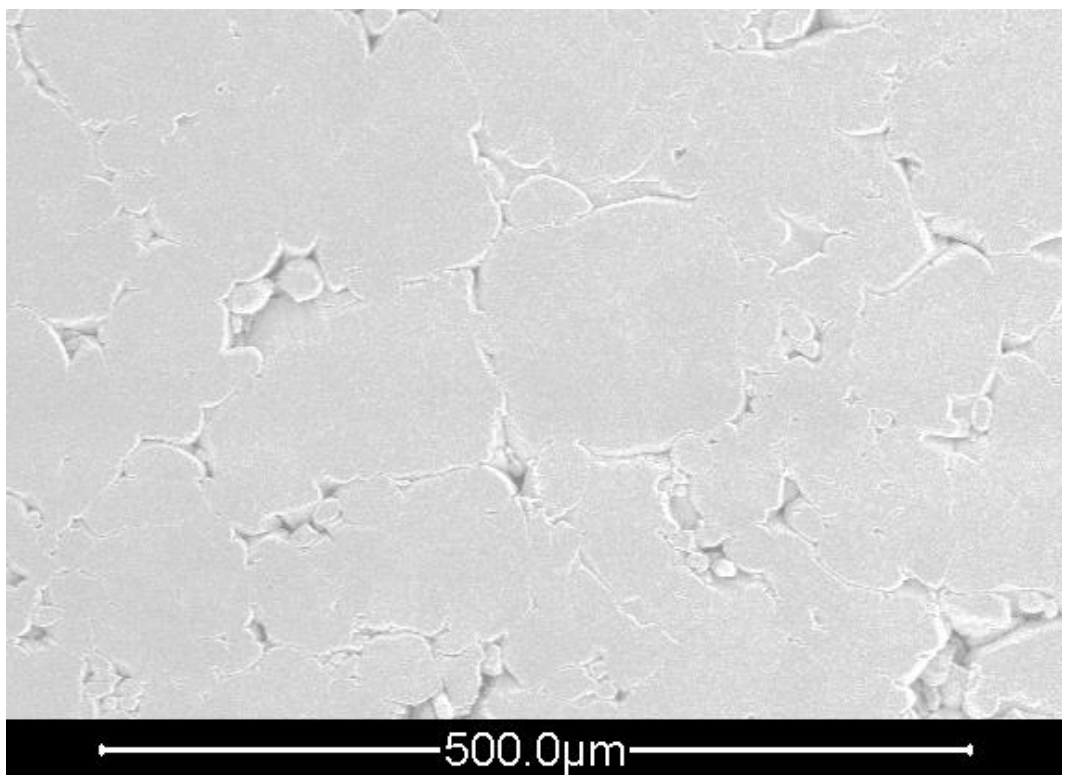  |
| **before exposure (zoom)** | 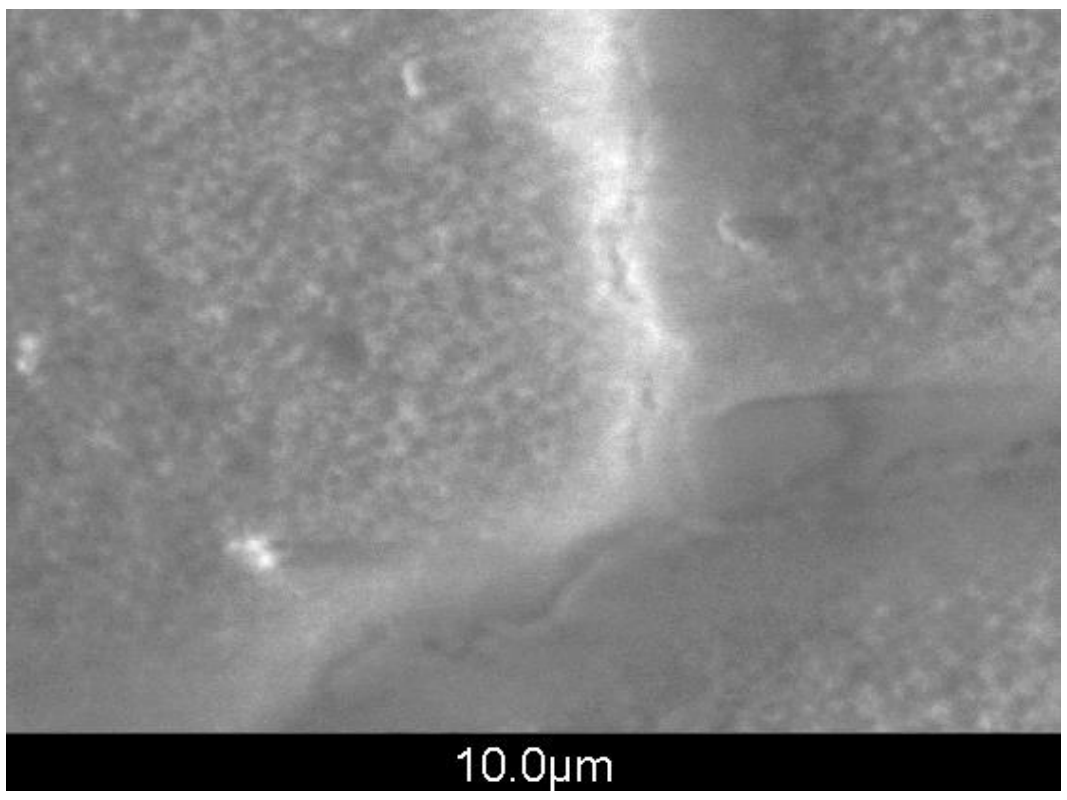  | 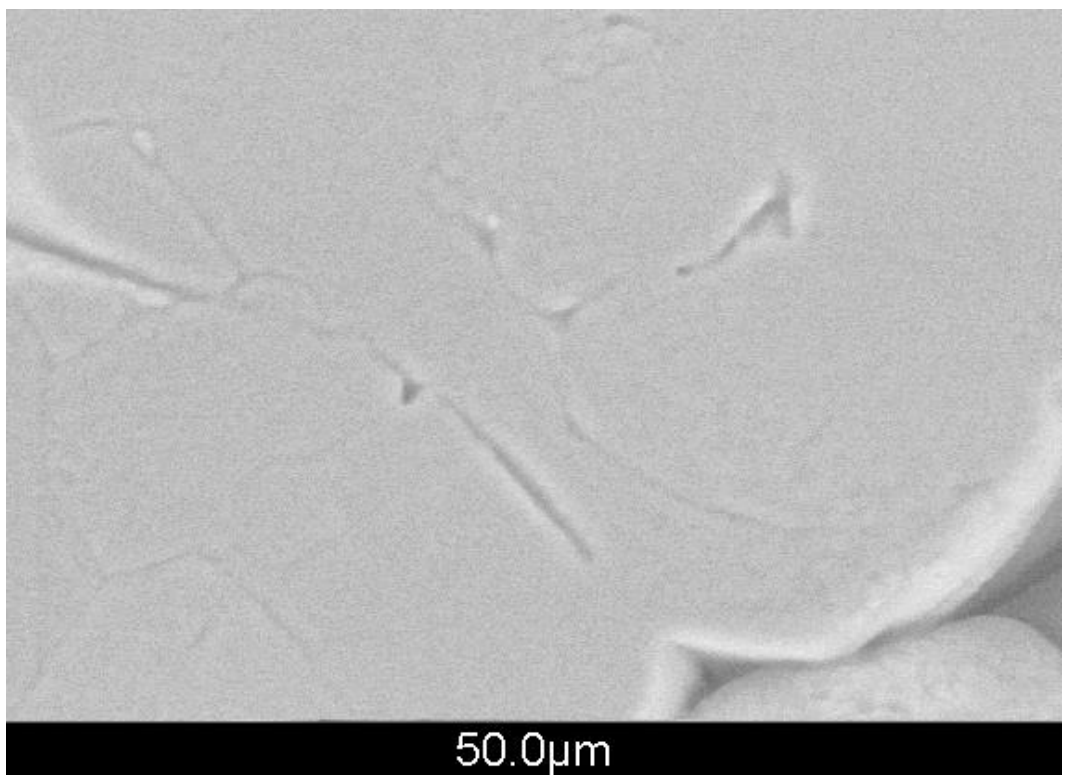  |
| **after exposure (4h)** | 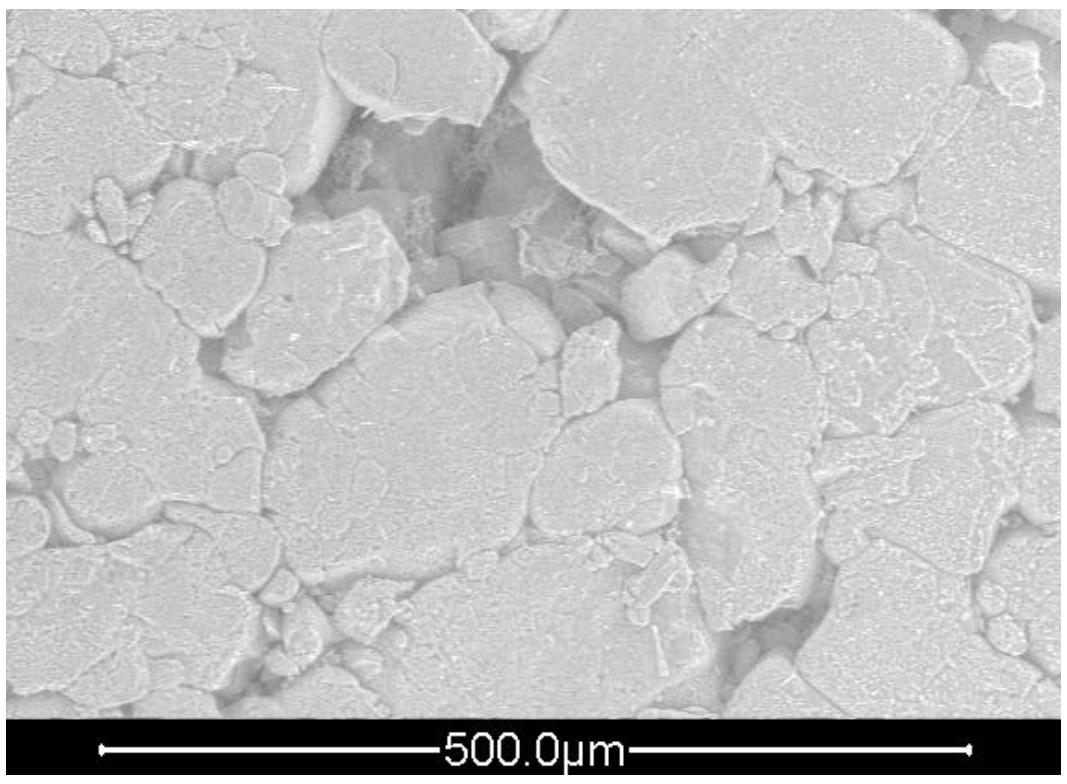  | 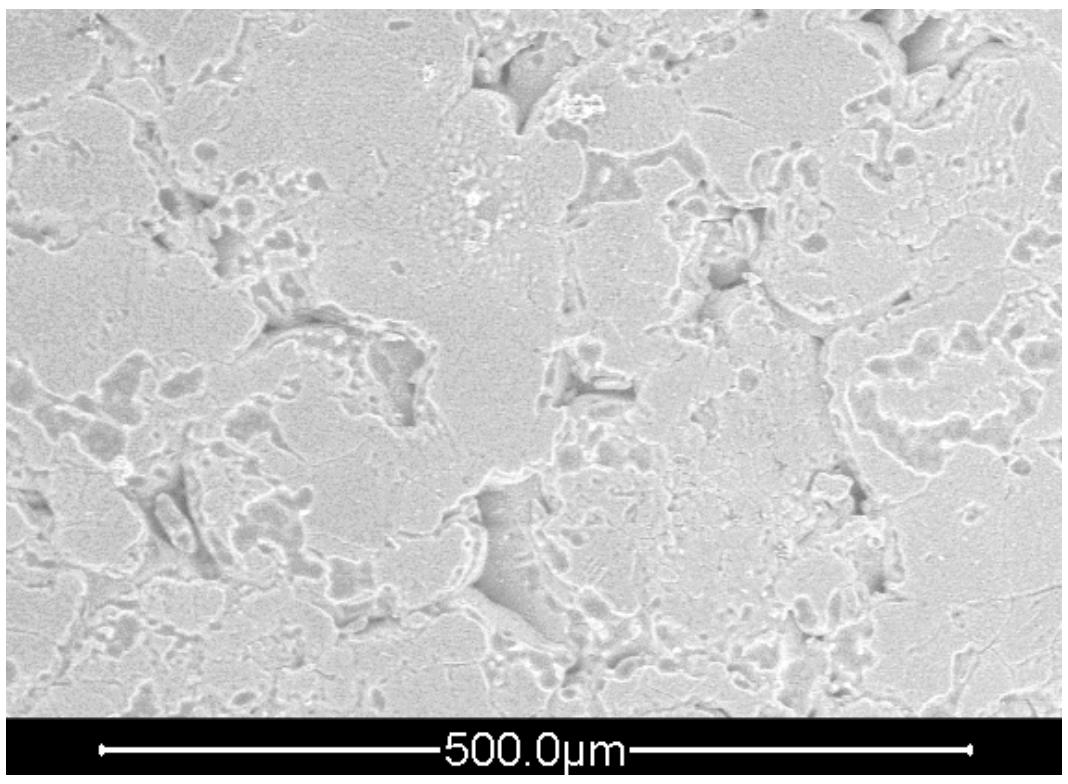  |

*Table 4: Surface of $P_{th}$ and $P_{kbr}$ before and after exposure seen by secondary electron detection with an environment SEM.*

Tholins spread on grids are more quickly altered by the exposure to plasma. The grid was cleared from tholins in less than 1h. SEM pictures were taken on pellet grids exposed during one hour to a $N_2$-$H_2$ plasma with 5% $H_2$ and at 1 mbar (see Table 5). Under plasma exposure, grains are eroded and become more porous. Holes of ~10 nm form inside the grains. After main grains are totally eroded, only remain small structures formed by particles of less than 100 nm. The sheath formed by the plasma in these conditions is ~100 µm. Therefore, erosion due to sputtering by ions should *a priori* lead to surface modifications with spatial resolution larger than the sheath size. Effects seen here are smaller than 100 µm, suggesting that they should be due to chemical processes.

This second technique has the advantage of exposing only tholins and not KBr grains to the plasma. However, when tholins are eroded, the structure is weakened. Packs of tholins are quickly removed from the mesh and the grid is cleared from tholins.

| | | |
|---|---|---|
| **before exposure** | 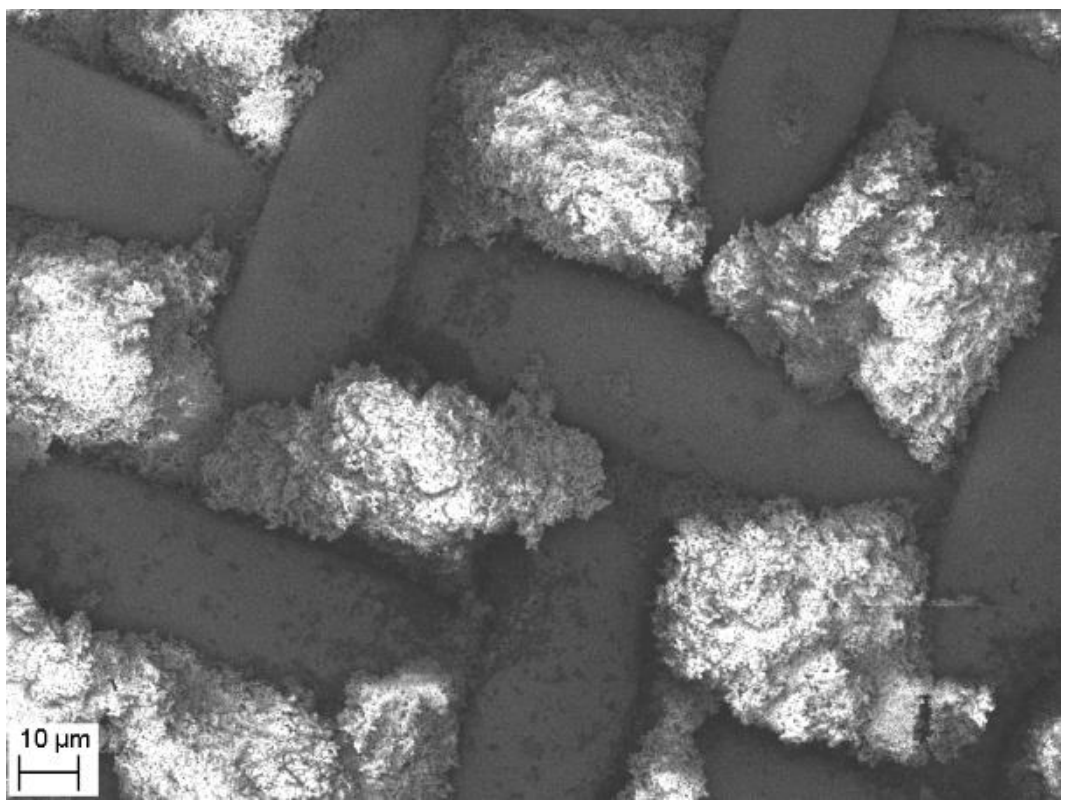 | 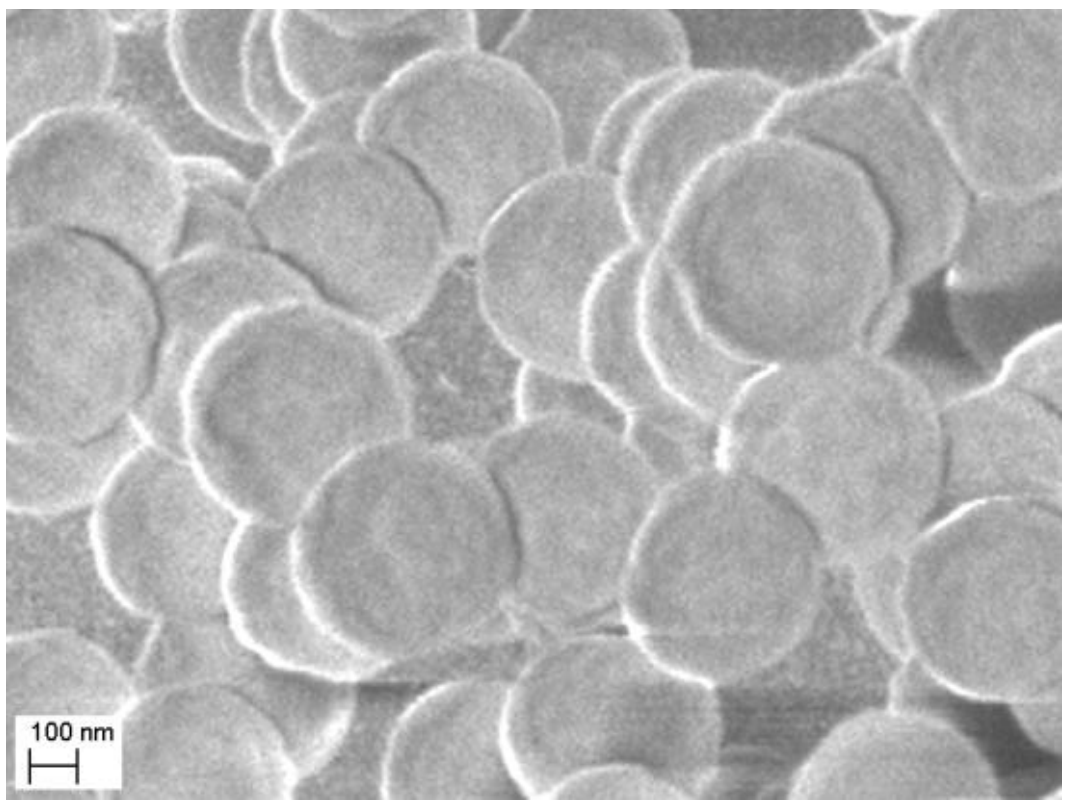 |
| **after exposure (1h)** | 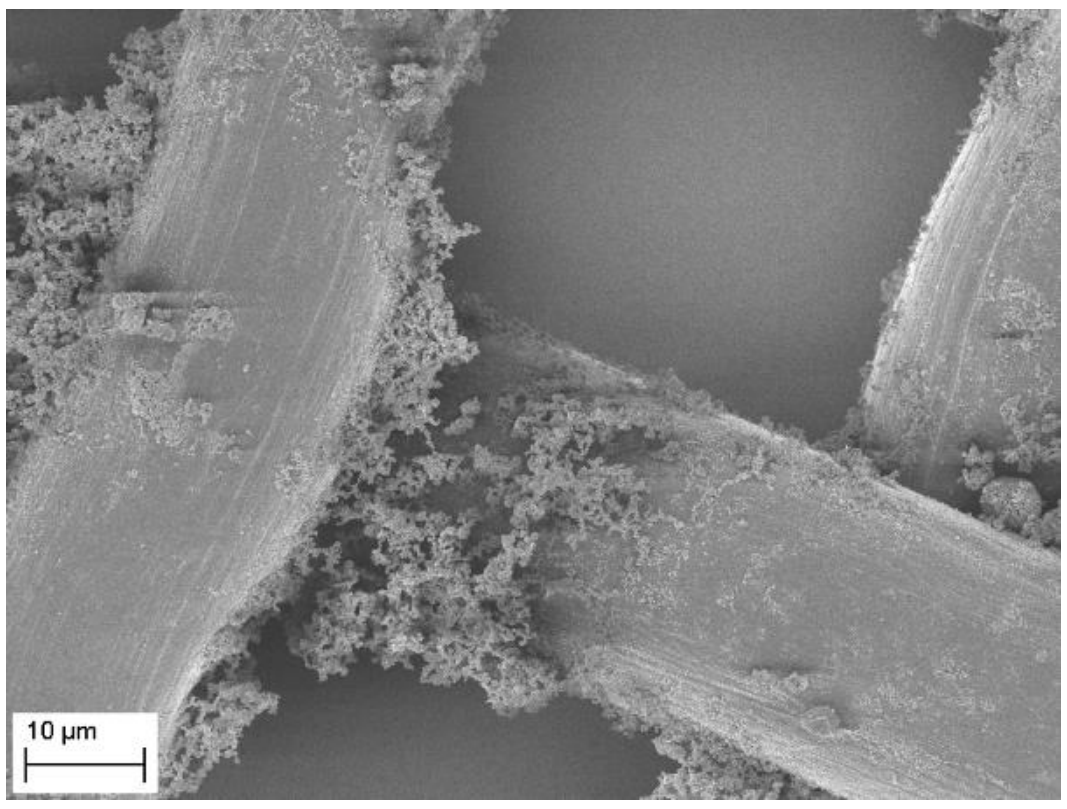 | 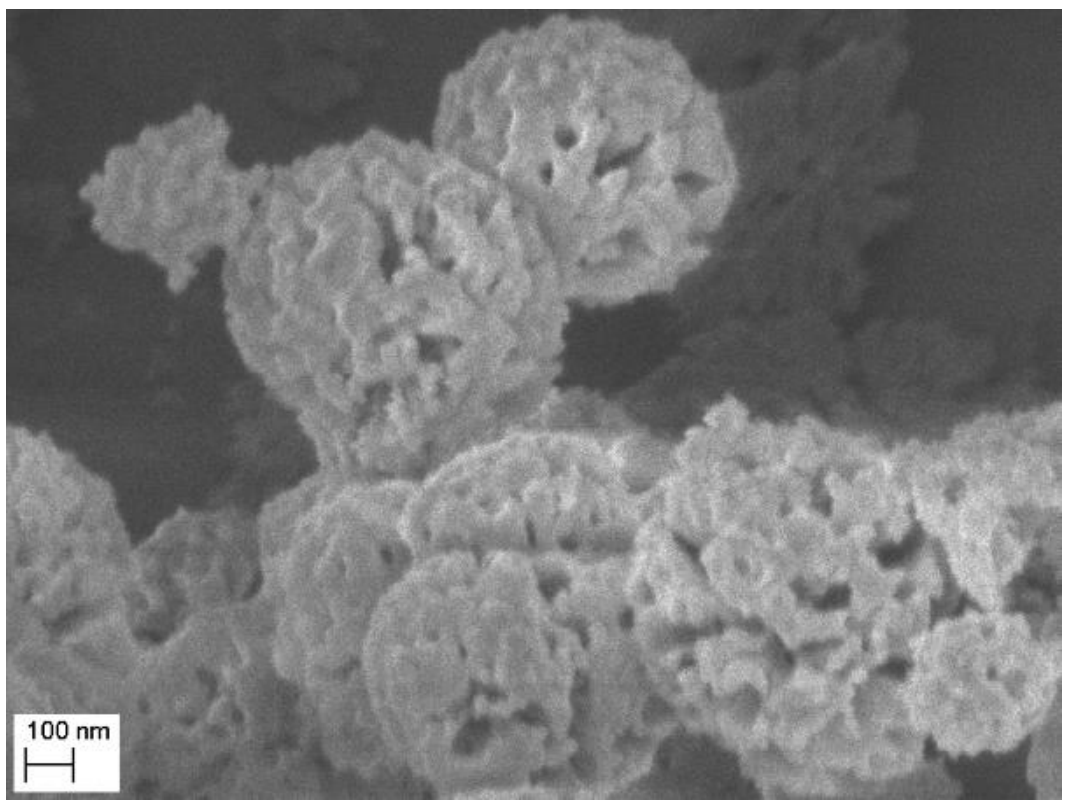 |
| **after exposure (1h)** | | 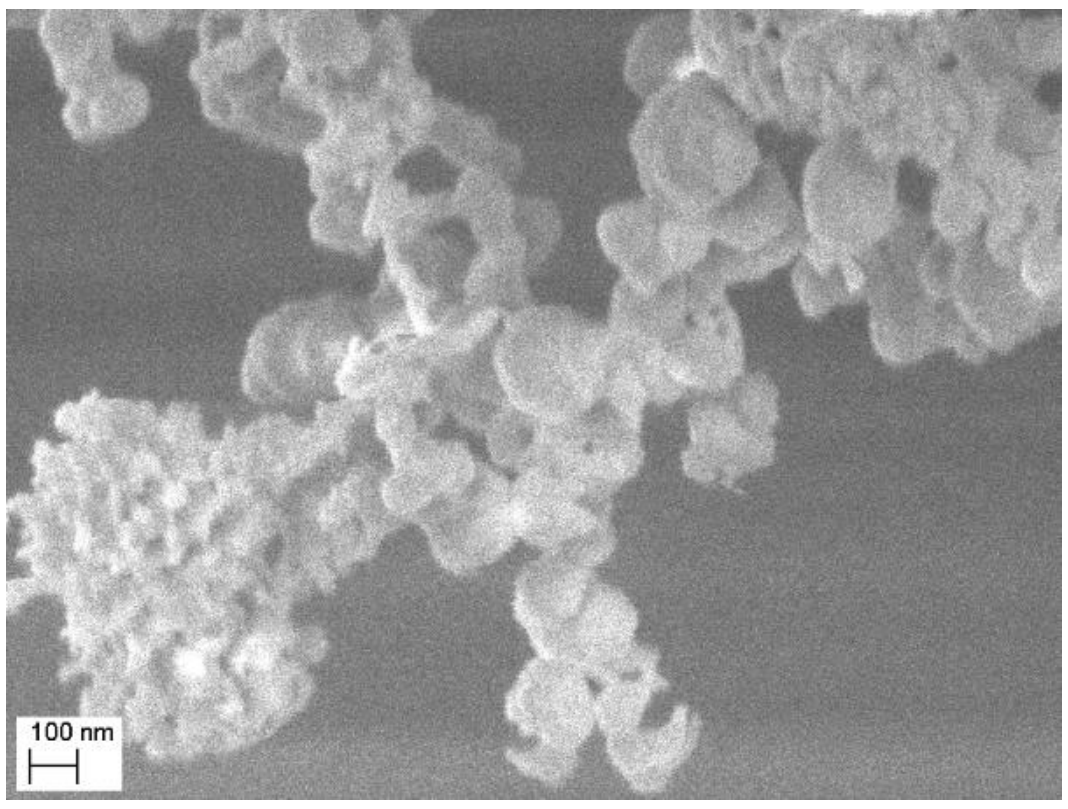 |

*Table 5: Surface of $P_{gr}$ before and after exposure with a FEG SEM.*

From the comparison of the SEM pictures for the two techniques of pellets, we conclude that in KBr-tholins pellets, KBr grains protect the tholins and slow down their evolution under plasma exposure. On the other side, grid pellets expose directly tholins to plasma, and meshes are quickly cleared by the plasma.

### *3.2.2 Pellet erosion monitored by IR spectroscopy*

The IR spectrum of pellets changes during exposure to plasma. Raw data gives the extinction, which is the combined effect of absorbance and diffusion:

$$extinction = -\log\left(\frac{I_{sample}}{I_0}\right)$$

$I_{sample}$ is the IR beam intensity collected after the tholins sample and $I_0$ is the intensity collected after a reference sample: pure KBr pellet non exposed to plasma or empty grid. Reflection is also generally included in extinction. However, here samples are positioned perpendicularly to the IR beam and they are not moved during the exposure. Therefore, the effect of reflection is supposed negligible for this study.

Grids decrease the transmission by a factor of 5, due to the physical obstacle of the threads and the enhanced diffusion through the meshes. Indeed, threads are ~38 µm large, which is close to the IR wavelengths (10 µm / 1000 $cm^{-1}$) [see Figure 11 in S.I. showing the IR intensity transmitted through pure KBr pellet and empty grid]. However, these effects are compensated by the division by $I_0$ in the computation of the absorbance. KBr pellets become rougher during the exposure to plasma (see section 3.2.1), and this increases the diffusion of IR light and therefore adds an offset to the extinction spectra during the experiment. To deduce the absorption spectra, the baseline is corrected by an adjusted polynomial [see raw data and polynomial fit in S.I. in Figure 12 and Figure 13].

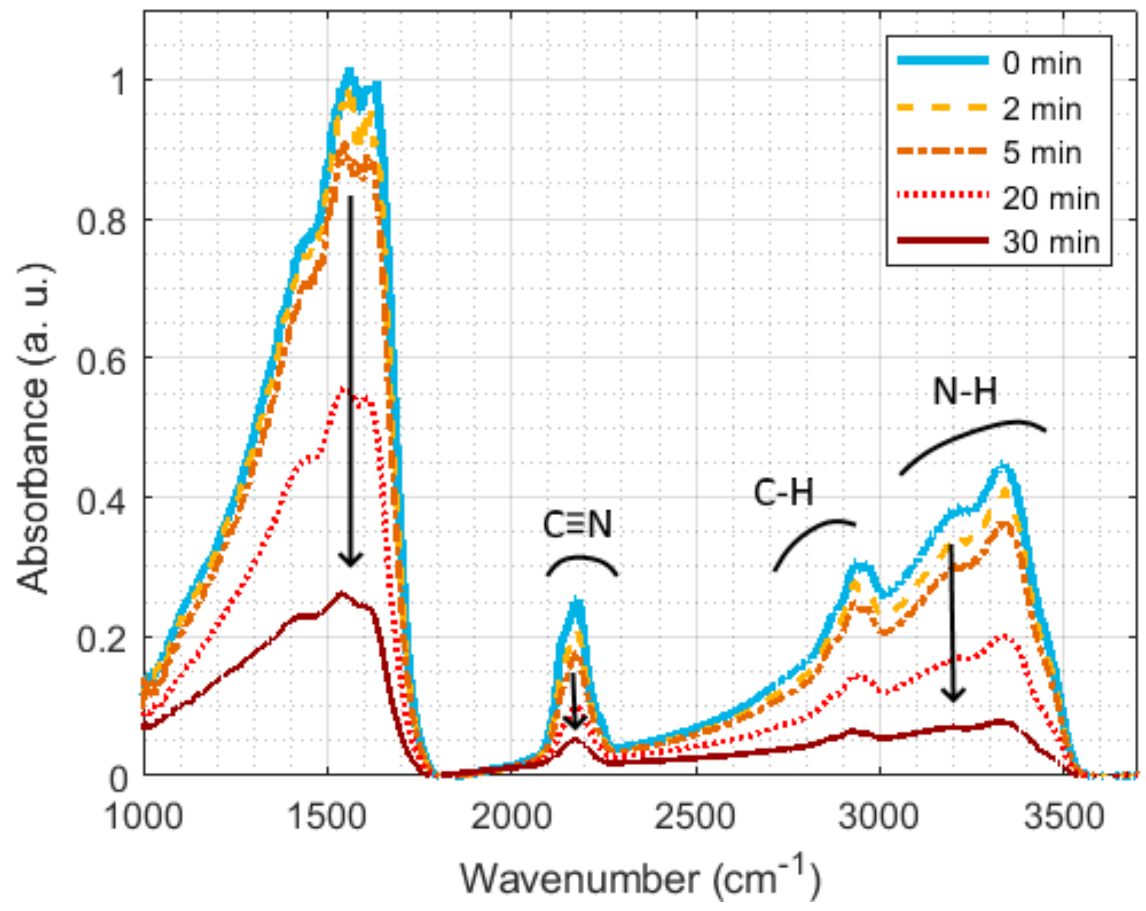

*Figure 5: Evolution of IR spectrum of tholins on grids ($P_{gr}$) during exposure to plasma.*

In both cases, the main effect is the global decrease of the absorbance (see Figure 5). It is consistent with the SEM observations described in section 3.2.1: some absorbent material is lost during the exposure to plasma.

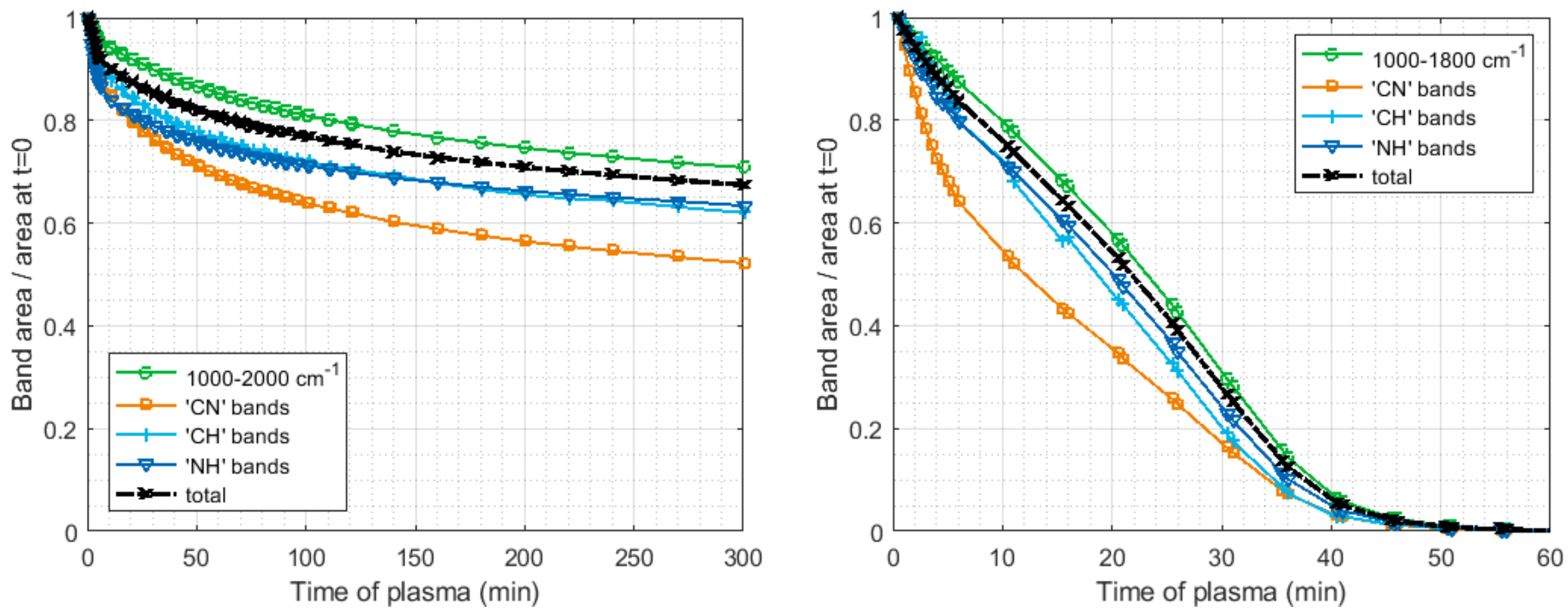

*Figure 6: Relative evolution of the three main absorption bands area during exposure to plasma: $P_{th}$ (left) et $P_{gr}$ (right). 'CN' (resp. 'CH') bands area is the area between 2020 and 2300 $cm^{-1}$ (resp. 2500 and 3100 $cm^{-1}$) once the underlying 'NH' band is removed. 'NH' bands area for $P_{th}$ (resp. $P_{gr}$) is taken between 2000 and 3700 $cm^{-1}$ (resp. 1800 and 3700 $cm^{-1}$) once the 'CN' and 'CH' bands are removed [see section 6.4 in S.I.]. The total area is taken between 1000 and 3700 $cm^{-1}$.*

The decrease of the bands area of spectra in time gives an idea of the loss time scale (see Figure 6). Results are different for the two pellets. The pellet with KBr loses 5% of its absorption in 5 minutes, but the decrease slows down after a few minutes and 70% of the absorbent material is still present after 3 hours. On the contrary, the decrease is linear for pellets with grids and all the absorbent material is gone after 45 min. This confirms the conclusions made from the SEM observations in section 3.2.1: KBr grains protect tholins from plasma erosion.

The areas of the major bands decrease slightly differently in time. In particular, CN bands disappear more quickly than the others. This suggests a different evolution of the chemical bonds and therefore the action of chemical processes during the erosion.

### *3.2.3 Conclusions on the comparison of KBr and grid pellets of tholins*

<table>
<tr><th></th><th>KBr pellet ($P_{th}$)</th><th>Grid ($P_{gr}$)</th></tr>
<tr><td>Advantages</td><td>Best method for IR analysis<br>- very good transmission in IR</td><td>Best method to expose aerosols to plasma<br>- direct exposure of tholins to plasma seen on SEM pictures</td></tr>
<tr><td rowspan="2">Drawbacks</td><td>Exposure aerosols/plasma hampered by KBr grains<br>- surface passivated by KBr grains (SEM pictures)<br>- slower decrease of IR absorbance during exposure to plasma</td><td rowspan="2">IR transmitted intensity 80% lower than with KBr pellets: lower signal to noise ratio.<br>- 75% absorption / reflection by threads<br>- diffusion through meshes of 38 µm and threads of 25 µm, close to IR wavelengths. Enhanced by the apparition of empty meshes after exposure to plasma.</td></tr>
<tr><td>Increased chemical pollution<br>- KBr is highly hydrophilic, it contains some water, which can interact / absorb on tholins during the preparation of the pellet.<br>- presence of an -OH band at 3500 $cm^{-1}$ and a C=O band at 1700 $cm^{-1}$, due to absorption of water and oxidation of the surface of tholins. Trapped in the KBr matrix, they are not entirely removed by heating.<br>- possible chemical interaction between $Br^-$ and positive ions formed in the discharge or derived from tholins.</td></tr>
</table>

*Table 6: Comparison of KBr and grid pellets: advantages and drawbacks.*

Both methods can be used to obtain IR spectra of tholins. However, each of them has its drawbacks (see Table 6): KBr pellets give the best IR spectra, but slow down the interaction of tholins with plasma and induce unwanted chemical effects. Therefore, it is fundamental to compare both techniques to conclude on the IR absorption of tholins.

## 3.3 Chemical modifications

### *3.3.1 Evolution of the spectra*

The relative evolution of bands on Figure 6 shows that nitriles and isonitriles bands are the most eroded by plasma. On the other hand, the NH and CH bands decrease at exactly the same speed. On Figure 3, films of tholins have an increase of CH bands compared to amines as what is seen with powders. Films are produced during a longer exposure to $N_2$-$CH_4$ plasma than powders. As no change is seen concerning relative evolution of CH and NH here, we conclude that the differences between films and powders are not due to long exposure to $N_2$-$H_2$ plasma species. Methane should play a fundamental role.

To study more closely the evolution of chemical functions of samples, each of the main broad bands is renormalized to its maximum. Interpretations on the evolution of the shapes of bands are therefore relative to these reference points.

Interesting modifications are observed on the nitrile and isonitrile bands (C≡N, see section 3.3.2). The decrease of unsaturated structures such as C=C and C=N is observed thanks to the disappearance of a band at 1665 $cm^{-1}$ (see section 3.3.3), and the decrease of specific C-H bands at 2950 $cm^{-1}$ (see section 3.3.4).

Excluding erosion, the amine band does not show any strong modification, but only a small decrease (< 4% in intensity compared to the maximum at 3340 $cm^{-1}$) of the left part of the band, below 3200 $cm^{-1}$ [presented in section 6.5 in S.I.]. This large band can be attributed to ammoniums or carboxylic acids. This loss also happens during heating. It can therefore be due to the loss of ammoniums which are highly unstable, or to the desorption of volatiles organic compounds (VOCs) formed at the surface of tholins by oxidation when they are exposed to air. The increase of diffusion with grids (from 1800 to 3300 $cm^{-1}$) and the loss of –OH bands (~3550 $cm^{-1}$) from water desorption on KBr pellets modify also this broad band. Nevertheless, we can conclude that amine bands do not evolve much during plasma exposure.

Bands showing modifications are discussed in the following parts. The baseline of each band is adjusted with a polynomial fit [explained in section 6.4 in S.I.] and bands are normalized at their maximum or at a selected reference point. Each plot of IR spectrum evolving in time goes along with a plot subtracting all spectra to the spectrum taken at t = 0. These plots enhance changes seen, and enable a clear view of the apparition and disappearance of bands. However, one should keep in mind that those curves depend on the reference normalizing point chosen.

### 3.3.2 Evolution of the nitrile and isonitrile bands

Bands between 2040 and 2260 $cm^{-1}$ are the most modified by the exposure to plasma. They are usually associated with nitriles and isonitriles functional groups which have strong stretching absorption modes in this range of wavenumbers. On the sample before exposure, the major band at 2177 $cm^{-1}$ is therefore usually associated with conjugated nitriles. At 2135 $cm^{-1}$ one can see the signature of isonitriles (R-N≡C), or maybe carbodiimines (R-N=C=N-R). The band at 2245 $cm^{-1}$ is due to saturated aliphatic nitriles (see Table 7).

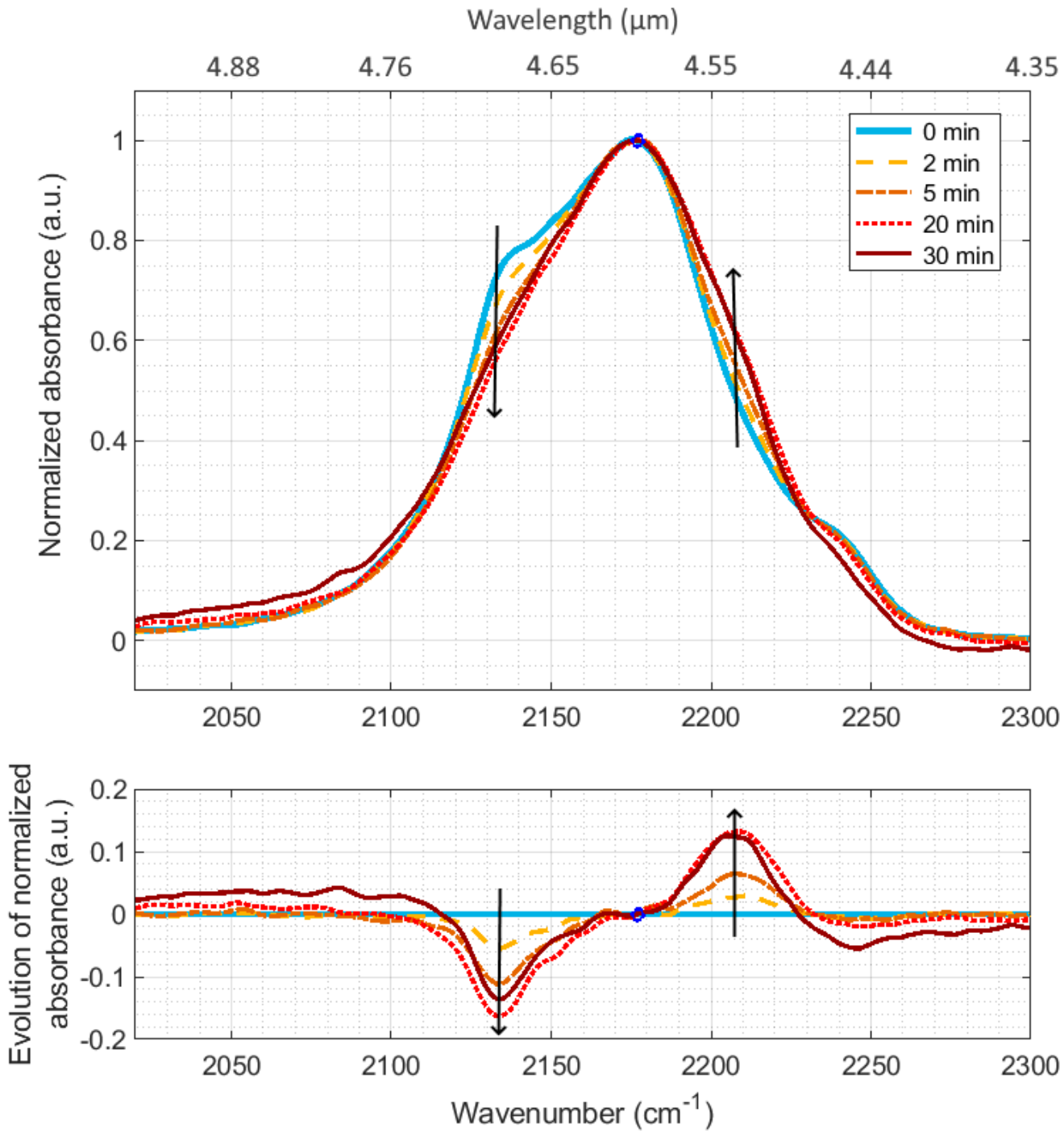

*Figure 7: Evolution of the IR nitrile band during exposure to plasma. Measurements with $P_{gr}$.*

| cm-1 | bond | evolution | reference |
|---|---|---|---|
| 2135 | isonitriles (R-N≡C)<br>and/or carbo-diimides (-N=C=N-) | quicker loss than the other bands | S04, I04 |
| 2177 | conjugated nitriles (R-C≡N)<br>and/or isonitriles with short 'R' | *reference point of the study* | S04, I04<br>T08, M99 |
| 2210 | β-unsaturated nitriles, aryl nitriles<br>and/or cyanamides (>N-C≡N) | increase compared to the other bands | S04, I04 |
| 2245 | saturated aliphatic nitriles | none | S04, I04 |

*Table 7: Band attributions in the range 2040-2260 cm$^{-1}$. All are stretching vibrations. S04 refers to* (Socrates, 2004), *I04 to* (Imanaka et al., 2004), *T08 to* (Truica-Marasescu and Wertheimer, 2008) *and M99 to* (Mutsukura and Akita, 1999)

When exposed to plasma we see similar evolutions for both pellets $P_{th}$ and $P_{gr}$. The isonitrile group at 2135 cm$^{-1}$ disappears more quickly than the other bands (cf Figure 7). In isonitriles, the N≡C bond is a terminal unsaturated and reactive function, which can easily be hydrogenated by $H_2$ and H in the plasma, explaining its rapid decay. This band can also be due to carbo-diimides (-N=C=N-) functions.

On the other side a new band of unsaturated nitriles grows around 2210 cm$^{-1}$. It could be associated to β-unsaturated nitriles and/or cyanamides (>N-C≡N), which have previously been detected in MS/MS measurements in the fragment $C_2N_3^-$ (Carrasco et al., 2009). There could be two explanations to this evolution: either this band resists more to plasma than the other nitriles, or it comes from the formation of additional functions, which could be linked to the transformation of isonitriles and/or carbodi-imides. Indeed, with grid pellets, bands at 2135 and 2210 cm$^{-1}$ evolve with similar timescales and intensities (~15% of the main band).

Alkynes (C≡C) absorb also at these wavelengths, at 2100-2150 cm$^{-1}$ if mono-substituted, and at 2190-2260 cm$^{-1}$ if di-substituted. However, these absorptions should have weak absorption bands compared to nitriles (Socrates, 2004), and previous GC-MS measurements on these tholins do not identify alkynes (Morisson et al., 2016). With KBr pellets the variations are slower and the band at 2210 $cm^{-1}$ is less intense.

Samples have also been analysed during heating at 80°C, at $10^{-6}$ mbar or under gas flux of $N_2$-$H_2$ at 4 mbar. In both cases, nitrile bands shift slightly during heating (with changes in intensity < 5% of the main band) but come back to their initial position at the end [spectra are presented in section 6.6 in S.I.]. The pellet temperature during heating is similar to its temperature during plasma exposure. Therefore, observations discussed above on the experiment under plasma exposure are not due to a heating effect, but rather to reactions with plasma species.

### *3.3.3 Modifications of unsaturated structures at 1665 $cm^{-1}$*

Absorption bands from 1000 to 1750 $cm^{-1}$ show also some modifications under plasma exposure. Both on grids and KBr pellets, a large band around 1665 $cm^{-1}$ decreases strongly. We therefore present only results on $P_{gr}$ here (see Figure 8).

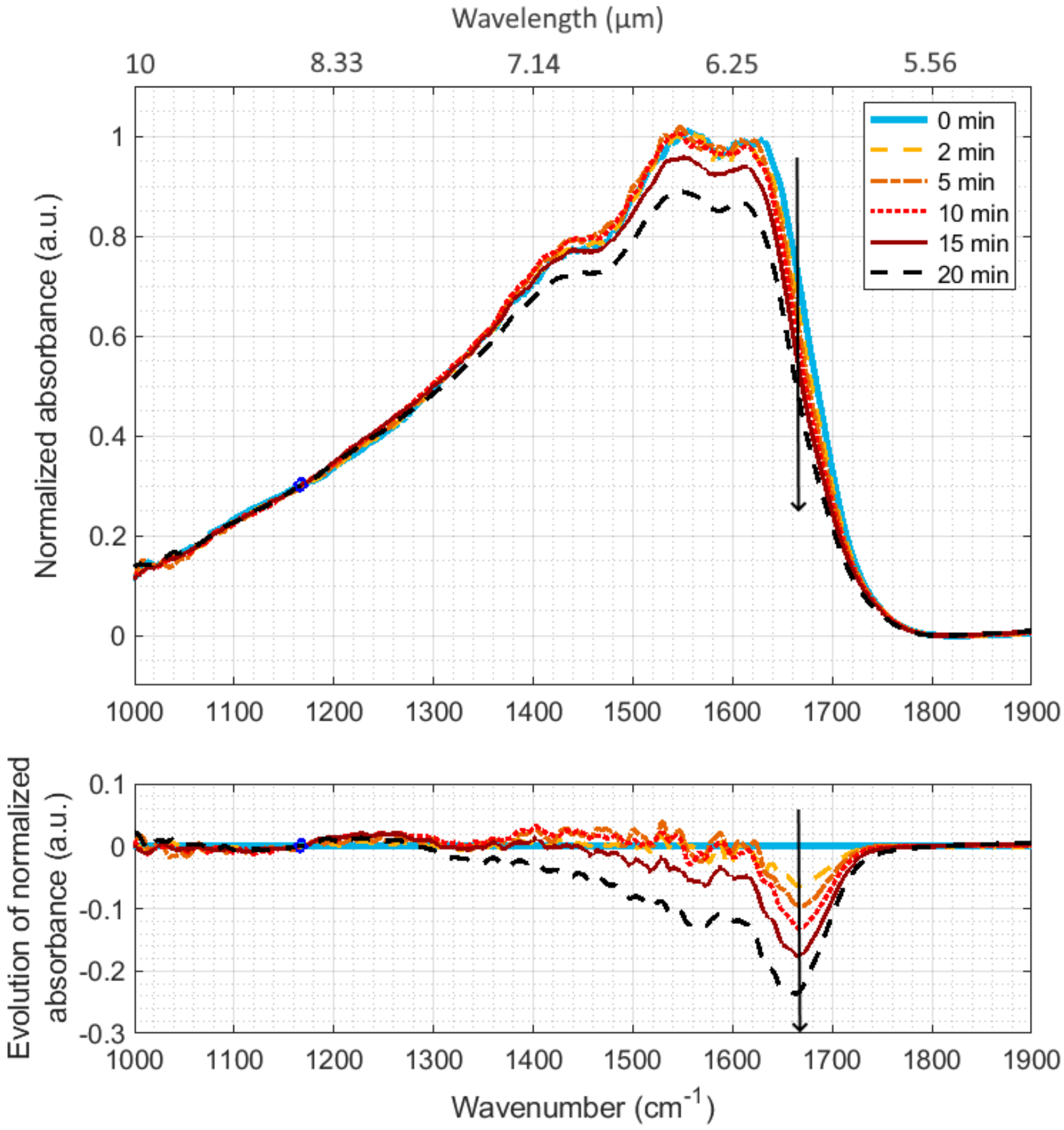

*Figure 8: Evolution of the IR band from 1000 to 1750 $cm^{-1}$ during exposure to plasma. Measurements with $P_{gr}$. The reference point for normalization was chosen at 1167 $cm^{-1}$ to give a clearer illustration of the evolution of the band. Indeed, the main peak at 1550 $cm^{-1}$ seems to evolve in time with reference to the rest of the band.*

| $cm^{-1}$ | bond | evolution | reference |
|---|---|---|---|
| 1000-1450 | various possibilities, especially bending vibrations of C-H bonds and stretching vibrations of C-N. | none | S04, I04 |
| 1550 | **N-H** (secondary or in C=N-H)<br>**N=N**<br>**C=N** (conjugated)<br>*pirimidine (C=C and C=N)* | small decrease at the end of the exposure | S04 |
| 1620 | **N-H** (primary)<br>**C=N** (unsaturated)<br>**C=C** (aromatic) | | S04, I04 |
| 1665 | R-**C**H=**N**-R'<br>**C=C** (especially tri- or tetra- substituted, or from >C=C-N<) | decrease under plasma | S04 |
| 1690 | **C=O** (from α,β-unsaturated or aryl aldehydes / ketones)<br>**C=C** (from α,β-unsaturated amines: $CH_2$=C-N<) | small decrease under heating | S04 |

*Table 8: Band attributions in the range 1000-1750 $cm^{-1}$: stretching vibrations of C=O, C=C, C=N and N=N, and deformation vibrations of C-H and N-H. S04 refers to* (Socrates, 2004) *and I04 to* (Imanaka et al., 2004).

Attributions in the wavenumber range 1000-1700 $cm^{-1}$ are ambiguous because several bonds have close absorption bands (see Table 8). In particular, bands at 1550 and 1620 $cm^{-1}$ can be due to secondary and primary amine (N-H) deformation vibrations, but also to unsaturated bonds such as C=C, C=N or N=N. The band at 1665 $cm^{-1}$ can be due to C=N stretching vibrations in R-CH=N-R' or to C=C stretching vibrations if tri- or tetra-substituted or in the >C=C-N< structure. The relative decrease of this band compared to the others (see Table 8 and Figure 8) shows that unsaturated bonds such as C=C and C=N are more easily modified by the $N_2$-$H_2$ plasma that the N-H and C-H bonds. This result is consistent as reactive $H_2$ and H species in the plasma will affect mainly unsaturations through hydrogenation reactions.

Before the exposure, during heating, only a thin band around 1690 $cm^{-1}$ decreases of 5% in intensity with reference to the maximum at 1550 $cm^{-1}$ [spectra presented in section 6.6 in S.I.]. Heating at ~100°C does not affect tholins formed in PAMRE (He et al., 2015). Therefore, the loss of the band at 1690 $cm^{-1}$ could be attributed to the desorption of Volatile Organic Compounds (VOCs) during the heating phase. Indeed, this band at 1690 $cm^{-1}$ could be attributed to stretching vibrations of C=O, or C=C in unsaturated amines (see Table 8). These VOCs should be mainly formed by oxidation processes at the surface of tholins when they are exposed to air (Pernot et al., 2010). On KBr grains, this band continues to decrease during plasma exposure: it seems all VOCs have not desorbed after the heating phase, especially the ones trapped inside the KBr grains pellet. However, thanks to experiments on grids, we are sure that modifications seen under plasma exposure at 1665 $cm^{-1}$ are not due to heating, but to plasma species.

### *3.3.4 Modification of unsaturated structures seen on CH bands*

Bands from 2700 to 3050 $cm^{-1}$ are due to C-H bonds. $CH_2$ and $CH_3$ structures have at least 2 different absorption bands, from symmetric and asymmetric stretching vibrations. Symmetric vibrations of acyclic $CH_2$ and CH, and of aliphatic $CH_3$ absorb at wavelengths inferior to 2900 $cm^{-1}$. Between 2900 and 3000 $cm^{-1}$, different bands superimpose: symmetric vibrations of acyclic $CH_2$ and CH, and aliphatic $CH_3$, plus symmetric and asymmetric vibrations from $CH_3$ bonded to unsaturated functions or >CHCN. Details are given in Table 9. No absorption band is observed at wavelengths superior to 3010 $cm^{-1}$, which discards the presence of C-H bonds on aromatic structures (Abdu et al., 2018; Socrates, 2004).

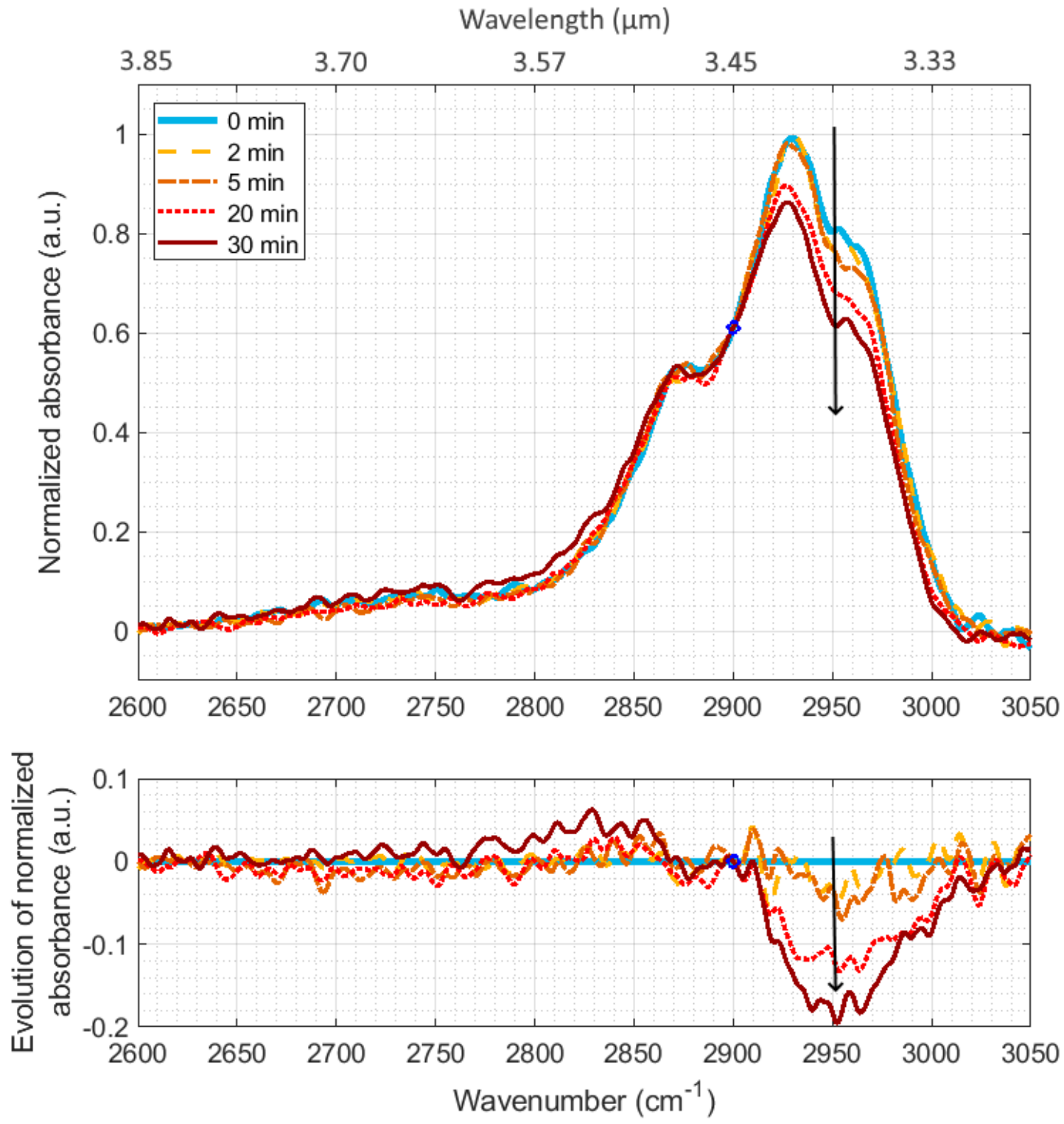

*Figure 9: Evolution of the IR band from 2600 to 3050 $cm^{-1}$ during exposure to plasma. Measurements with $P_{gr}$. The reference point for normalization was chosen at 2900 $cm^{-1}$ to give a clearer illustration of the evolution of the band. Indeed, the main peak at 2930 $cm^{-1}$ seems to evolve in time with reference to the rest of the band.*

| $cm^{-1}$ | bond | evolution | reference |
|---|---|---|---|
| 2650-2800 | -N-$CH_3$ or –N-$CH_2$-(?) (sym) | none | S04, I04 |
| 2850 | acyclic $CH_2$ (sym) | none | S04, D18, A18 |
| 2875 | aliphatic $CH_3$ (sym) | none | S04, I04, D18 |
| 2880-2890 | acyclic CH (weak) | none | S04, G12 |
| 2880-3010 | unsaturated –$CH_3$, 3 bands in this range | loss during plasma | S04 |
| 2900-2910 | aliphatic $CH_2$ (Fermi resonance) | | D05, D18 |
| 2930 | acyclic $CH_2$ (asym) | | S04, I04, G12, D18, A18 |
| 2935-3035 | -$CH_3$ linked to >CH-C≡N or –C(sat group)$_3$, 2 bands | | S04 |
| 2960-2970 | $CH_3$ (asym) | | S04, I04, G12, D18, A18 |

*Table 9: Band attributions in the range 2600-3050 $cm^{-1}$: stretching vibrations of C-H. S04 refers to* (Socrates, 2004), *I04 to* (Imanaka et al., 2004), *G12 to* (Gautier et al., 2012), *D05 to* (Dartois et al., 2005), *D18 to* (Dionnet et al., 2018) *and A18 to* (Abdu et al., 2018).

Exposure to plasma leads to the loss of a large band from ~2910 to ~3020 $cm^{-1}$ with reference to the rest of the bands (see Figure 9). As symmetric and asymmetric absorption bands should evolve similarly, the disappearance of this band should be due to the loss of structures with methyl bonded to unsaturated structures or to >CHCN. This agrees with the previous conclusions showing that C=C, C=N, and C≡N tend to disappear more quickly than saturated structures when exposed to a $N_2$-$H_2$ plasma.

On grids, no evolution is observed during heating, which confirms that the observations are due to plasma species interaction [see section 6.6 in S.I.].

However, on KBr pellets, other side effects are seen: it seems CH bonded to aldehydes are observed at 2700 and 2775 $cm^{-1}$ and decrease during exposure to plasma. This point is consistent with the decrease of C=O bands during exposure of KBr pellets in section 3.3.3. Besides, some changes are also seen during heating of pure KBr pellets, they could be due to water or VOCs still trapped between KBr grains. Anyway, as these effects are not seen on grid pellets, they are not due to tholins and we will not further analyse them.

## 3.4 Experimental conclusions

### 3.4.1 Comparison of KBr and grid pellets

In the present study we used two different kinds of pellets. The comparison of both gave us insights on their advantages and drawbacks. It also gave us confidence that the presented effects are related to tholins evolution under plasma exposure, as they can be discriminated from possible artefacts induced by KBr or metal grids. Indeed, all evolutions presented above are seen using both techniques. The ones seen only with one technique were immediately discarded and attributed to the influence of the technique used to make the pellet (KBr grains or metallic thin grids).

Concerning the comparison of the two methods, it turned out that grid pellets were more appropriate for our application. KBr pellets are often used for FTIR study of solids (Abdu et al., 2018), because it gives spectra with a really good signal to noise ratio, contrarily to grid pellets that transmit only 20% of the incoming light because of absorption, reflexion and diffusion [the transmitted intensity is presented in section 6.1 in S.I.]. However, in our case, the exposure of tholins to plasma is hampered by the presence of KBr. KBr grains physically protect tholins inside the pellet from plasma (cf results in section 2.3.1). They also trap water and VOCs inside the pellet and hinder them from desorbing during the cleaning heating before the exposure to plasma (cf results in section 3.3). This last point is enhanced by the fact that KBr is highly hydrophilic.

In conclusion, to avoid physical or chemical actions of KBr that could modify IR spectra, we recommend the use of grid pellets in this experimental configuration.

### 3.4.2 Effect of temperature in the plasma discharge

A difference between our experiment and Titan's atmosphere is the gas temperature which is not controlled at 170K in our experiment. To work at ambient temperature can affect the reaction rates, the products and their stability in the plasma. It should mainly accelerate processes. However, no further study on this subject can be done with this setup as condensation appears on the walls at low temperature. Nevertheless, it is necessary to check that a higher temperature does not induce by itself new chemical evolutions of the sample. Previous works (Bonnet et al., 2015; He et al., 2015; Morisson et al., 2016) studied the degradation of tholins under heating. Effects are seen only above 300°C, where tholins start to carbonize and create polyaromatic structures.

During our experiment here, the lightning of the discharge at 20 mA increases the gas temperature to about 60-80°C (Loureiro and Ricard, 1993; Pintassilgo and Guerra, 2017). According to the works cited in the previous paragraph, the temperature rise here should not degrade by itself the tholin chemical structure.

Measurements during heating are consistent with this hypothesis. They are described in S.I [see section 6.6]. Measurements of the temperature of the glass in contact with the pellet gives a temperature of ~80°C. No interesting evolution has been detected in IR analysis, just the desorption of water and VOCs, especially in the case of KBr pellets.

In conclusion, the physical and chemical modifications seen on the pellets are not due to a degradation by the rise of temperature but to the physical and chemical erosion by the $N_2$-$H_2$ plasma.

# 4. Discussion

## 4.1 Evolution processes in the plasma

Several evolution processes can be deduced from the results obtained here. First ion sputtering should be one physical process leading to the global erosion of the sample. Ions from $N_2$-$H_2$ plasma, and in particular hydrogenated ions, should be energetic enough to erode surfaces exposed to the plasma (Sharma et al., 2006; Tanarro et al., 2007).

Besides, we observed that the nanometric erosion of individual grains, under the dimensions of the plasma sheath linked to sputtering, and the evolution of IR bands reflect also chemical evolution processes (see section 3.3). The accelerated decrease of nitrile and isonitrile bands compared to the others shows a preferential erosion of these highly unsaturated bonds. On the other side, N-H and C-H bands evolve similarly and cannot explain the differences seen between powder and film spectra (see Figure 3). The study of bands around 1665 and 2950 $cm^{-1}$ seems to indicate a preferential erosion of unsaturated centres like C=N and C=C compared to simple bonds as N-H and C-H. In conclusion, we observed that unsaturations are more affected by plasma exposure. This is consistent with the fact that hydrogenated species in the $N_2$-$H_2$ plasma react with unsaturated sites.

The evolution of the nitrile and isonitrile band is simple enough to infer the chemical process at stake. The quasi-simultaneous evolution of two bands at 2135 and 2210 $cm^{-1}$ suggests that isonitriles (-N≡C) are converted into nitriles (-C≡N). These evolutions can also be explained by the transformation of carbo-diimides (-N=C=N-) into cyanamides (>N-C≡N). In both cases, more stable bonds are formed.

## 4.2 Modifications due to photon irradiation

On Titan, the upper atmosphere is ionized by energetic electrons coming from Saturn's magnetosphere, but also by solar photons. Especially, UV irradiation has a major impact in the upper layers (Gronoff et al., 2009; P. Lavvas et al., 2011), and among them VUV photons ionize easily nitrogen and methane in the atmosphere.

Laboratory plasmas also emit VUV photons, thanks to excited species. Consequently, surfaces exposed to plasma undergo combined effects of plasma particles and VUV photons. In particular, excited hydrogen and nitrogen emit strongly in the region 170-120 nm (Wertheimer et al., 1999). A study on the plasma VUV photons impact on polymer surfaces showed they can photochemically ablate the polymers and create volatile fragments (Hong et al., 2002). Therefore, this effect could appear in the experiment described here. However, further studies are required to quantify it. Besides, we expect VUV photon fluxes in the plasma to be weak compared to fluxes on Titan because of strong self re-absorption in the gas phase in the plasma reactor.

A recent study focused on the only effect of VUV photons on tholin films (Carrasco et al., 2018). To be consistent with the UV irradiation dose received by aerosols on Titan, samples were exposed during 24h to strong fluxes, using synchrotron light. It showed the de-hydrogenation of the chemical functions.

Results here do not include de-hydrogenation but other chemical modifications. Consequently, modifications observed in this work cannot be attributed to VUV photons. In our plasma discharge photons are therefore less efficient to evolve tholins than ions, electrons or radicals.

Consequently, at least two erosive agents could be at work on Titan's aerosols: VUV photons for de-hydrogenation and $N_2$-$H_2$ plasma particles that modify the unsaturated functions (see the above discussion in section 4.1). In the ionosphere, carbon growth and hydrogenation processes should be in competition, from the influence of carbon and hydrogen of plasma species derived from $CH_4$. However, just below the ionosphere, ion processes should stop and modifications should be dominated by VUV irradiation, leading to the de-hydrogenation of the organic grains.

## 4.3 Modifications due to atomic hydrogen irradiation

Similarly to VUV photons, atomic hydrogen counts among the reactive species present in Titan's upper atmosphere and likely to affect aerosols through heterogeneous reactions. Atomic hydrogen is mainly formed around 800 km on Titan due to direct photodissociation of methane and $CH_3$ and reaction of CH with methane (Lebonnois et al., 2003; Vuitton et al., 2019). It can reach a density up to $10^8$ $cm^{-3}$. Another source of atomic hydrogen is present lower, in the stratosphere, where unsaturated hydrocarbons undergo photodissociation.

The effect of atomic hydrogen on Titan's aerosols has been studied in the conditions of Titan's stratosphere and mesophere (below 600 km) by Sekine et al., 2008a. They observed a strong loss of H by hydrogenation at the surface of tholins and by recombination in $H_2$. No significant chemical evolution of the tholins has been noted, the main conclusion being the strong consumption (~60%) of atomic hydrogen by heterogeneous reactions at these altitudes (Sekine et al., 2008b; Vuitton et al., 2019).

Atomic hydrogen is also present in large quantities (~0.1%) in both the ionosphere of Titan (Vuitton et al., 2019) and in $N_2$-$H_2$ laboratory plasmas (Carrasco et al., 2013; Sode et al., 2015). Therefore, hydrogenation of tholins and recombination due to atomic hydrogen could also appear, both in the experiment presented here and in the ionosphere of Titan. As atomic hydrogen is not likely to chemically evolve tholins (Sekine et al., 2008a), modifications seen in the work presented here should be due to other plasma species, in particular ions.

## 4.4 Evolution expected on Titan

As the experiment described here aims to mimic phenomena happening on Titan, one should expect that physical and chemical modifications seen in the laboratory should also appear on Titan's aerosols. The experimental simulation simplifies the conditions in Titan's ionosphere. Only $N_2$-$H_2$ plasma species are present in the experiment while at least methane and VUV photons also play simultaneously a great role in Titan's ionospheric chemistry. Therefore, the processes studied here will be concomitant with other evolution processes on Titan.

Few information is known about aerosols in Titan's ionosphere, making it difficult to compare with experimental results. In the whole infrared range studied here, only the small part around CH bands can be compared to Cassini data. In addition, the ionosphere is not dense enough for infrared measurement and these spectra have been taken around 200 km, in the stratosphere (Kim et al., 2011).

VIMS spectra show a complex CH band group, with $CH_2$ and $CH_3$ signatures, in agreement with tholins here (see section 3.3.4). However, the main difference is the presence of absorption peaks above 3000 $cm^{-1}$ in VIMS data that are not seen with PAMPRE tholins. These bands are attributed to CH linked to aromatic structures (Socrates, 2004). Therefore, we showed that aromatic structures are *a priori* not formed by $N_2$-$H_2$ plasma processes. However, as VIMS spectra are taken far below the ionosphere, many other processes could be thought of to explain this formation, with for instance a large abundance of polycyclic aromatic hydrocarbons (PAHs) in the atmosphere (López-Puertas et al., 2013) or the condensation of ice on aerosols (Kim et al., 2011).

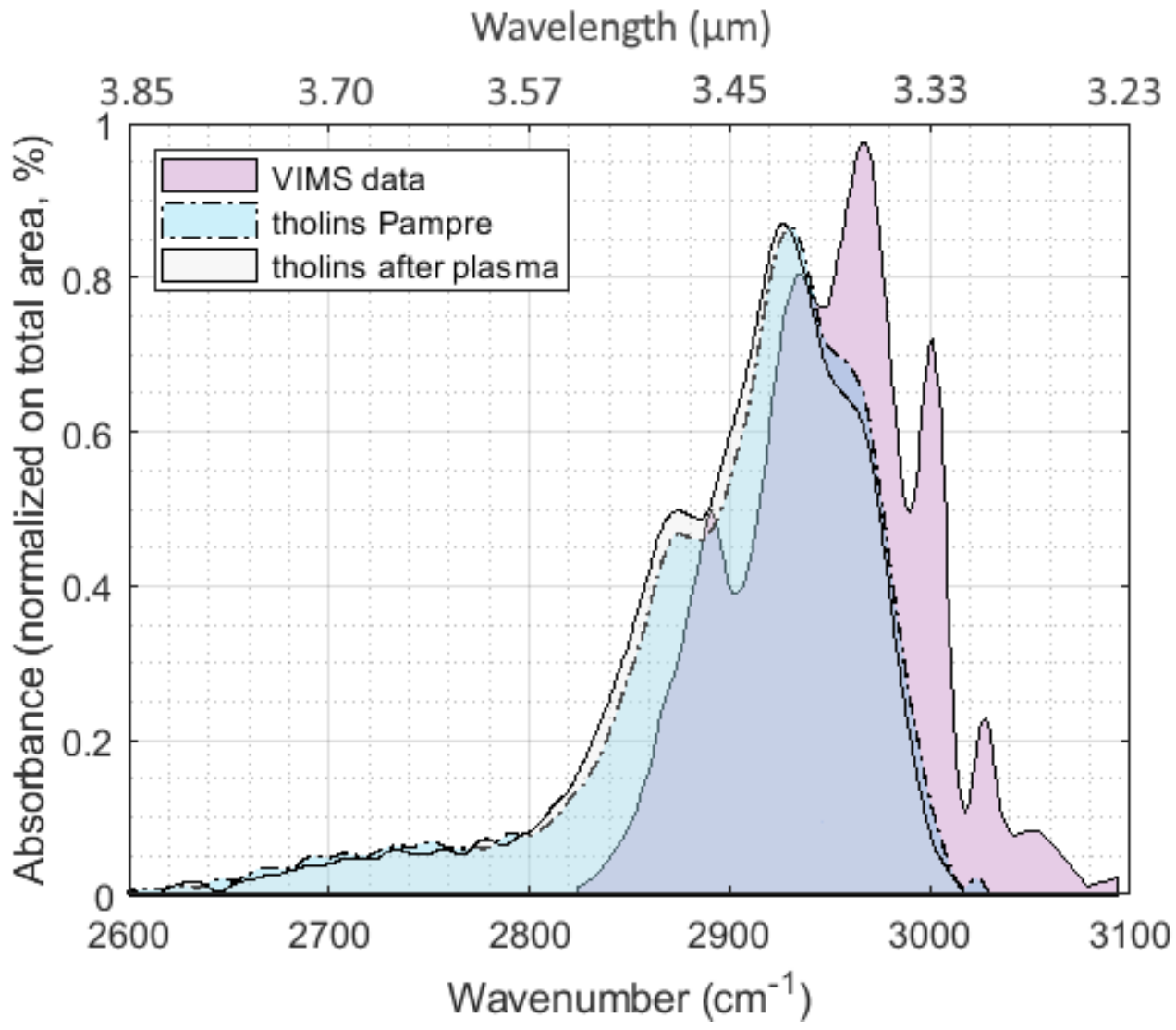

*Figure 10: Comparison of infrared CH absorption bands in our experiment partially mimicking Titan's ionosphere and VIMS data in the stratosphere* (Kim et al., 2011).

We showed that the exposure of grains to $N_2$-$H_2$ plasma species leads to the formation of 10-nm erosion structures on grains (in section 3.2.1). This effect is certainly enhanced by the absence of methane in the mixture that would tend to form new carbon chains around. However, it shows that nano-structures could be formed and increase the sticking coefficient of grains that would tend to absorb smaller molecules or form agglomerates of grains, as suggested by (P Lavvas et al., 2011). This enhanced granularity should also influence the efficiency of organic aerosols to act as condensation nucleus lower in the atmosphere of Titan.

# 5. Conclusion

The experimental simulation presented here aims to mimic the evolution of Titan's aerosols in its ionosphere. Thanks to analogues of these aerosols and to a plasma discharge in nitrogen with 1% of hydrogen, we showed that organic matter typical to Titan's ionosphere is sensitive to its plasma environment. In particular, originally round smooth grains are eroded by $N_2$-$H_2$ plasma species, leading to nano-structures at their surface. Chemical functions are also modified. Isonitriles and/or carbo-diimides decrease strongly, contrarily to more stable functions as nitriles and/or cyanamides which grow compared to the global nitrile band. Double bonds as C=C and C=N are more affected by the plasma exposure than amines and C-H bonds. On the contrary, N-H and C-H absorption bands keep a similar ratio in intensity and their shape does not vary. VIMS spectra in Titan's stratosphere show new aromatic CH bands compared to laboratory aerosols mimicking ionospheric particles, which are not explained by the plasma erosion studied there. This suggests that other processes also modify aerosols in Titan's atmosphere.

Concerning the erosion due to $N_2$-$H_2$ plasma, the next steps are to understand the role of each of the plasma species. Indeed, the gas phase interacting with the aerosols is composed by neutral, electrons, ions and excited species. Besides, the influence of hydrogen can especially arouse interest as protons $H^+$ and small protonated ions are mobile and very reactive.

On Titan, aerosols evolve during their sedimentation through the ionosphere and to the surface. Their physical and chemical properties are certainly modified by the various environment encountered, and we showed $N_2$-$H_2$ plasma species in the ionosphere should play a role in it.

# 6. Supplementary information

## 6.1 IR intensity through KBr pellet and grid

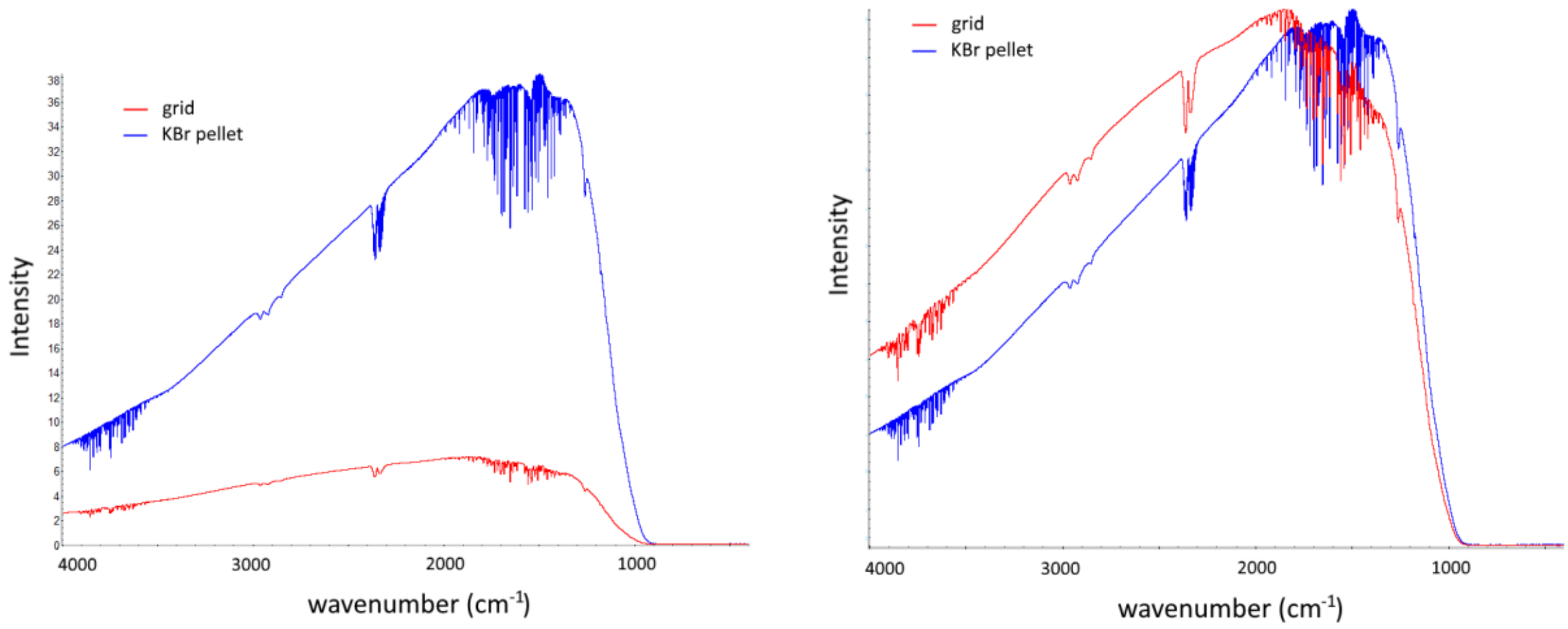

*Figure 11: IR intensity through reference pellets (without tholins). Normalized on the right panel.*

KBr is transparent in IR wavelengths. It is often used to make pellets mixed with a material to study. On the other hand, grids of 38 µm absorb 80% of IR light and give noisier spectra. Grids also deform the IR lamp emission spectrum by diffraction through the meshes.

## 6.2 Evolution of the baseline during the exposure

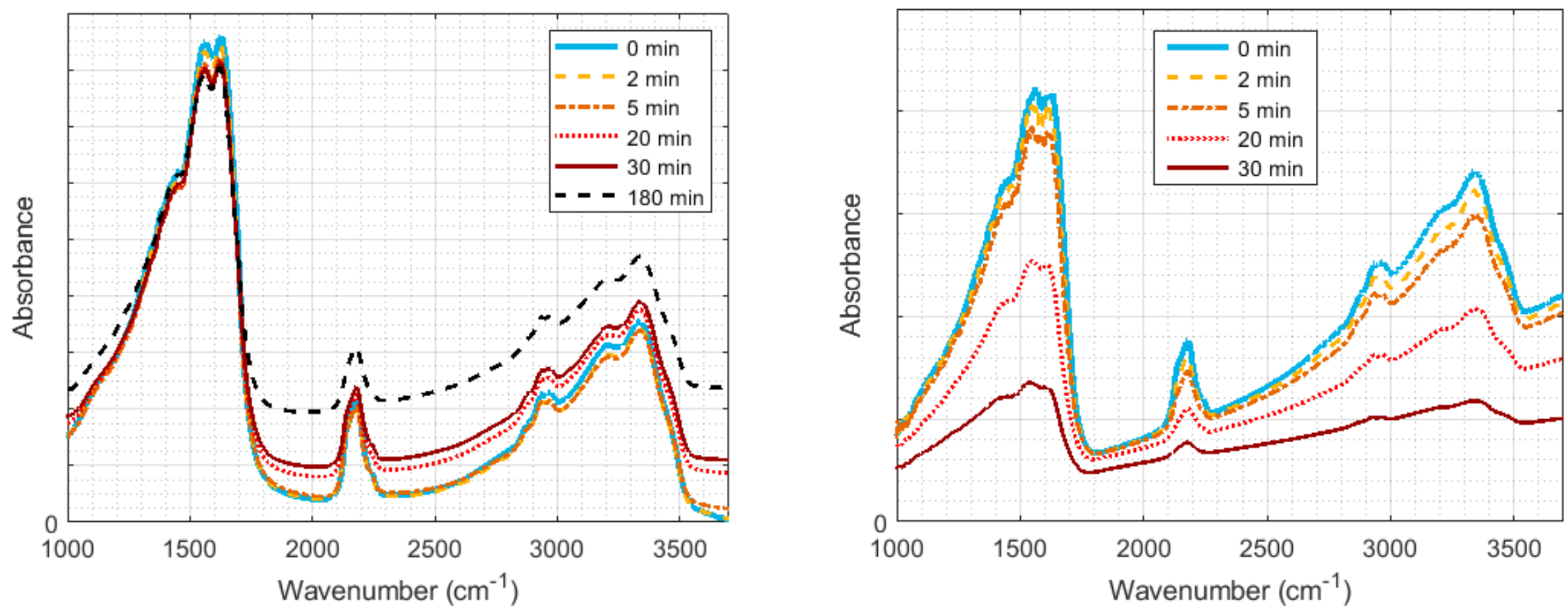

*Figure 12: Evolution of IR spectra of samples during exposure to plasma, raw absorbance. Left: on a KBr pellet ($P_{th}$). Right: on a grid pellet ($P_{gr}$).*

Spectra evolve during exposure to plasma. Concerning the KBr pellet, the mean value of the baseline increases and can be explained by the enhancement of diffusion through pellets as they become rougher. In parallel, for both KBr and grid pellets, the total amplitude of the extinction decreases, which mainly means a loss of absorbing particles. Both phenomena are consistent with the visual and SEM observations, detailed in section 3.2.1.

## 6.3 Correction of the baseline

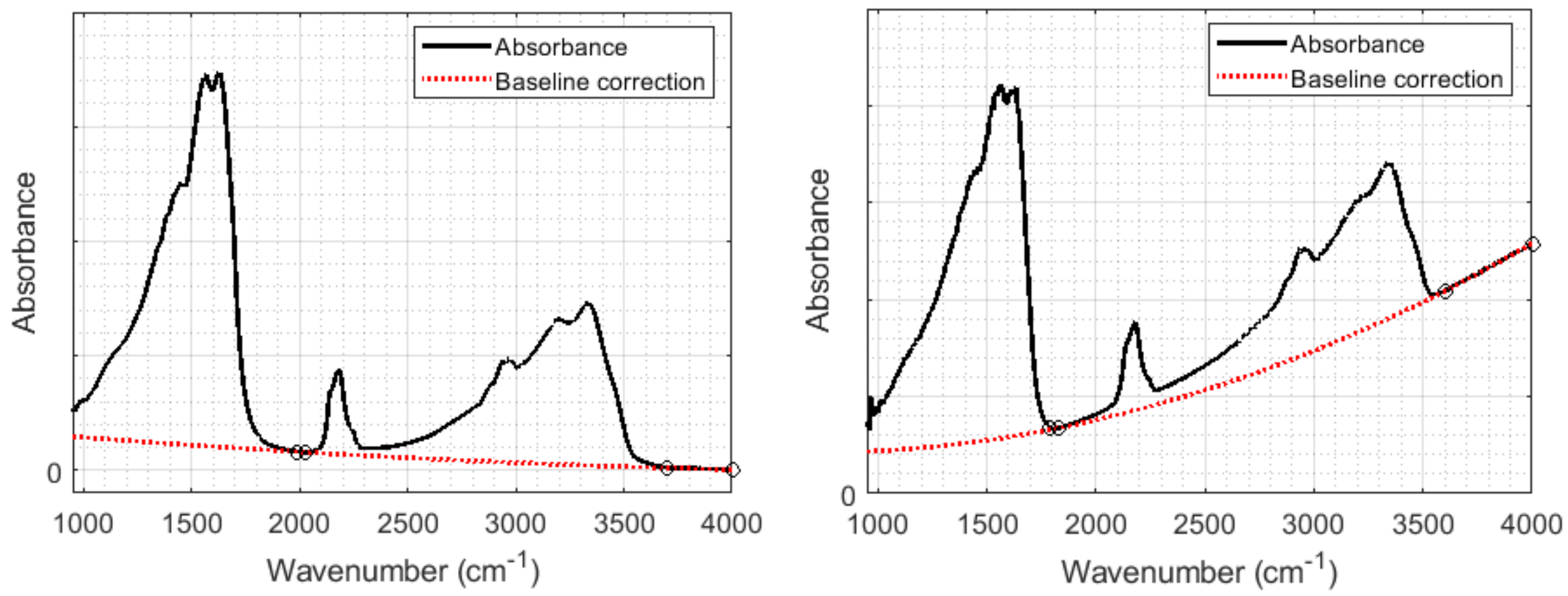

*Figure 13: Polynomial correction of the baseline of absorbance spectra of tholins. Left: on a KBr pellet ($P_{th}$). Right: on a grid pellet ($P_{gr}$).*

The baseline can be corrected to remove the major effect of diffusion. Here a polynomial subtracted to the original baseline. It is defined by two reference wavelength ranges where tholins do not absorb (from 1790 to 1830 cm$^{-1}$ and from 3600 to 4000 cm$^{-1}$ in the case of grids; from 1990 to 2030 cm$^{-1}$ and from 3700 to 4000 cm$^{-1}$ in the case of KBr pellets). The polynomial fit is computed for each spectrum.

## 6.4 Areas of 'CN', 'CH' and 'NH' bands

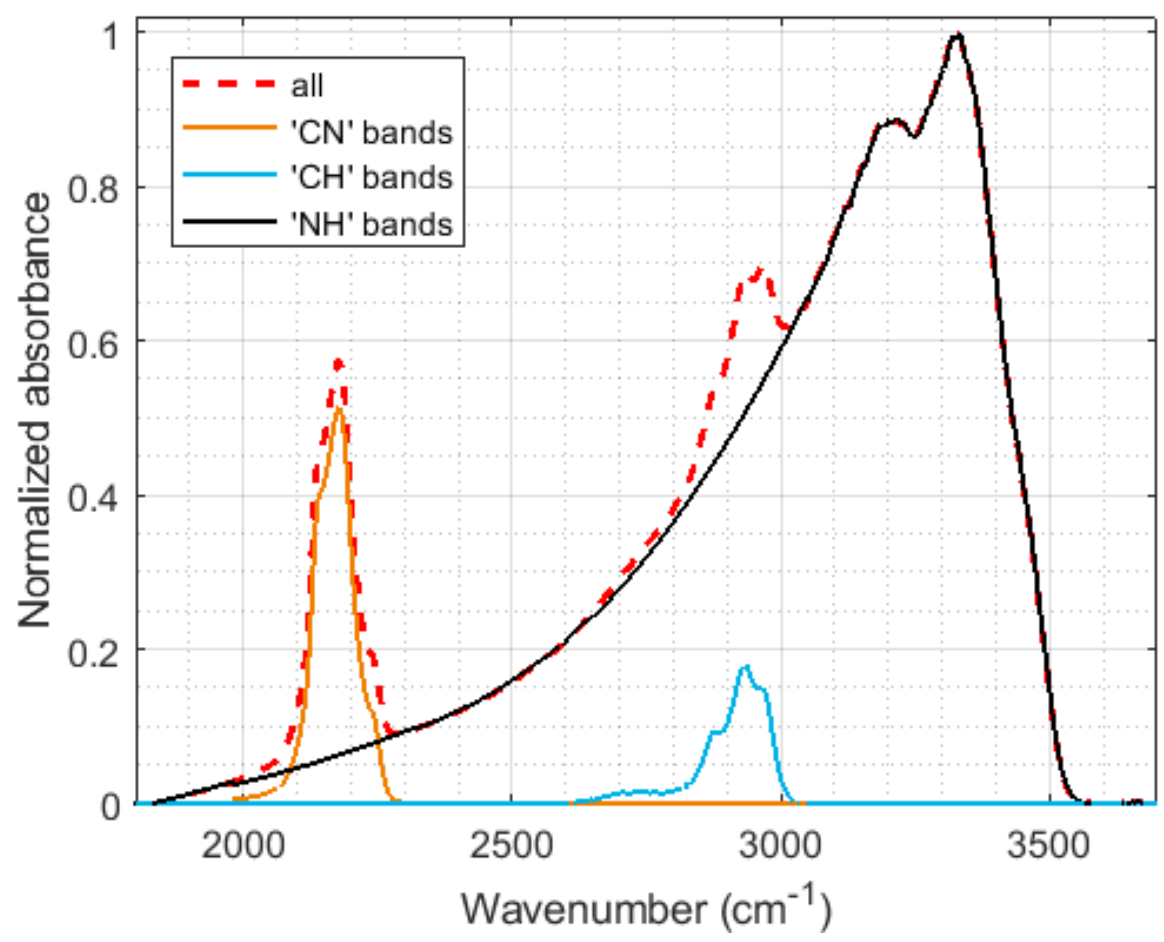

*Figure 14: 'CN' and 'CH' bands are dissociated from the 'NH' bands thanks to polynomial fits. IR spectrum of $P_{gr}$ before exposure to plasma.*

NH (and OH) bands (1800-3600 cm$^{-1}$) are superimposed with CN (2000-2300 cm$^{-1}$) and CH bands (2500-3100 cm$^{-1}$). We removed the CN and CH bands thanks to polynomial fits on the left wing of the NH bands.

## 6.5 Only minor modifications on the amine band

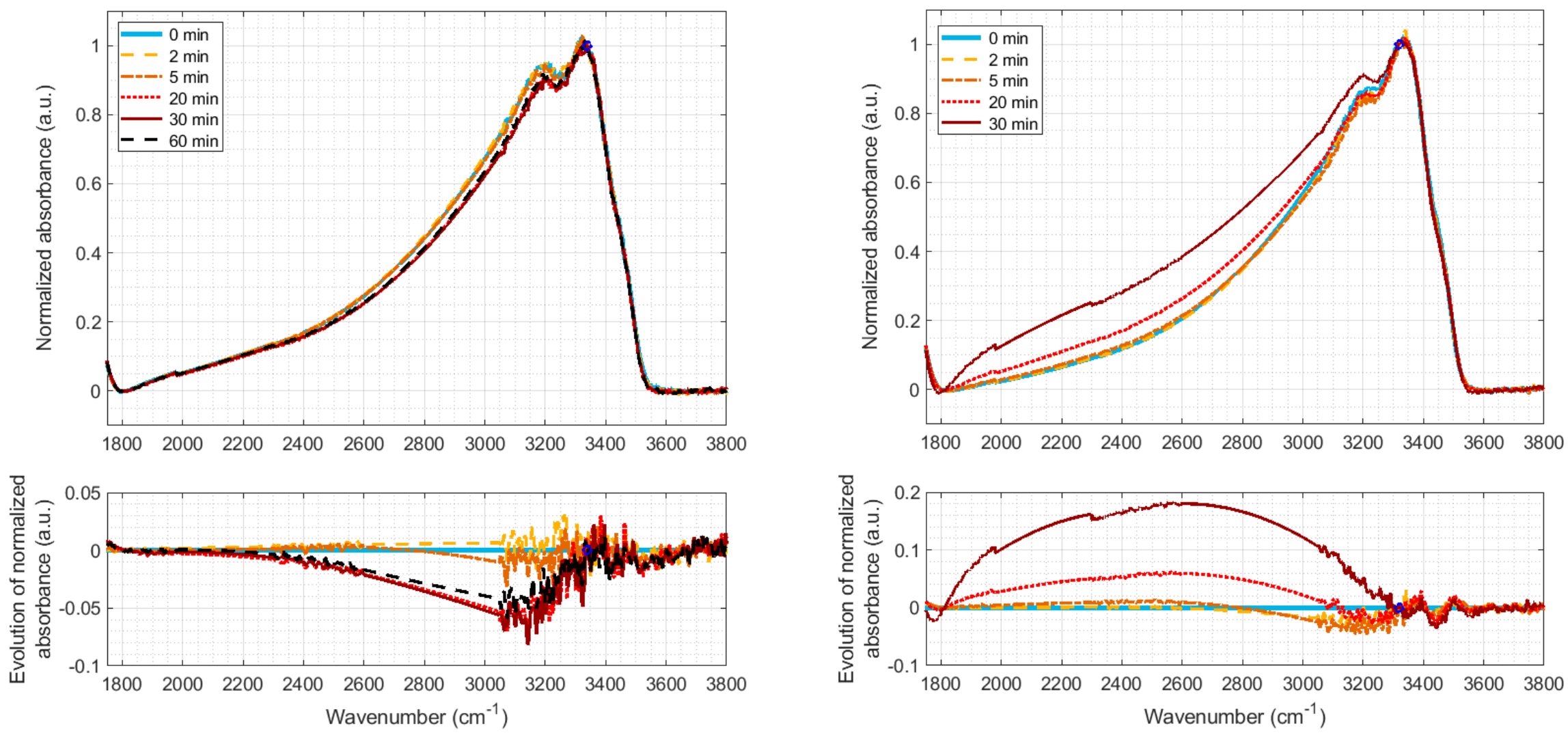

*Figure 15: Evolution of the IR band from 1800 to 3600 cm$^{-1}$ during exposure to plasma, after the removal of 'CN' and 'CH' bands. Measurements with $P_{gr}$. The reference point for normalization was chosen on the main peak at 3338 cm$^{-1}$. Left: evolution during heating. Right: evolution under plasma exposure.*

No strong modification happens in the amine band concerning tholins. A small decrease of a large band around 3200 cm$^{-1}$ could be due to the loss of ammoniums under heating. The large band from 1800 to 3200 cm$^{-1}$ appears at the end of plasma exposure, but only on grid samples. It should be due to diffusion through the emptying meshes. KBr pellets show a strong loss at 3550 cm$^{-1}$ (-OH bonds) due to the desorption of water.

## 6.6 Evolution of tholins during heating

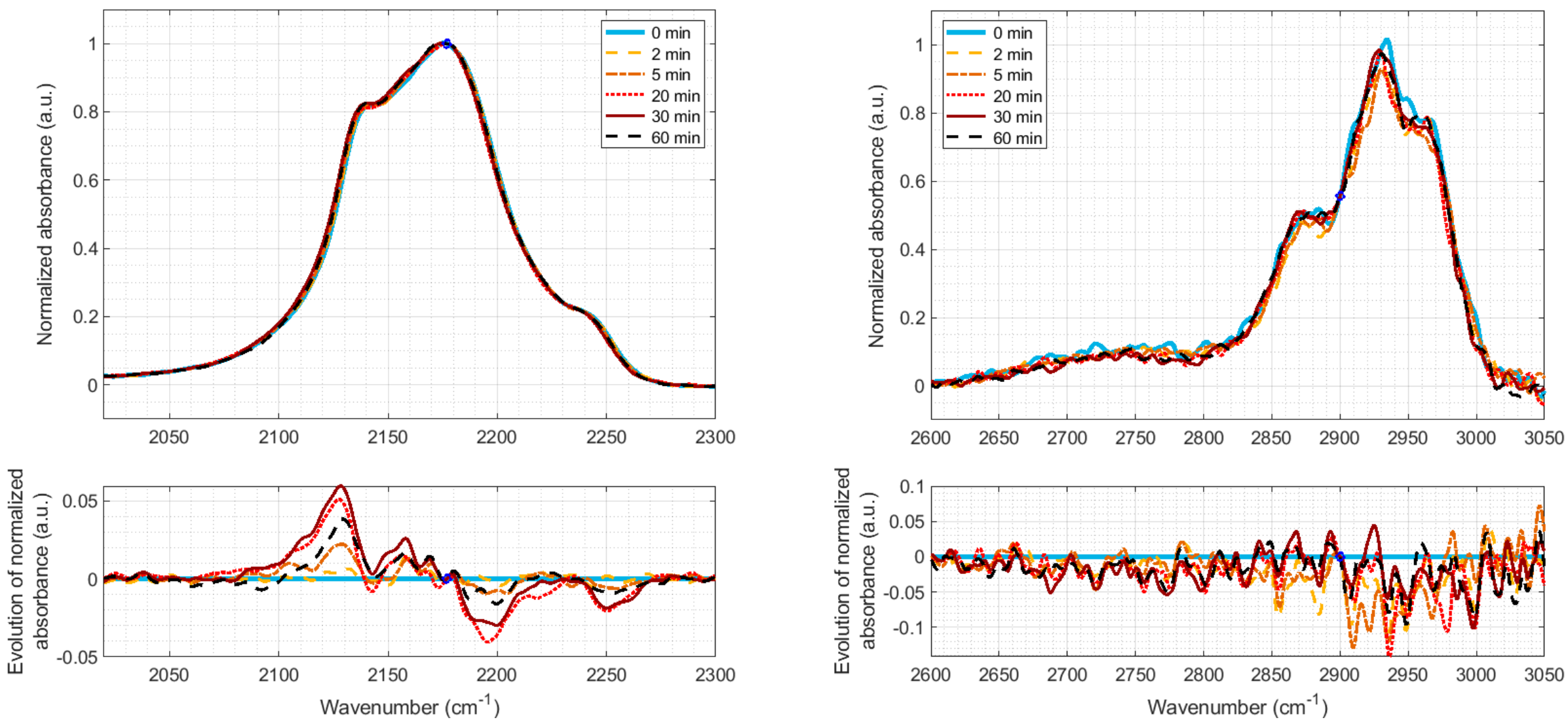

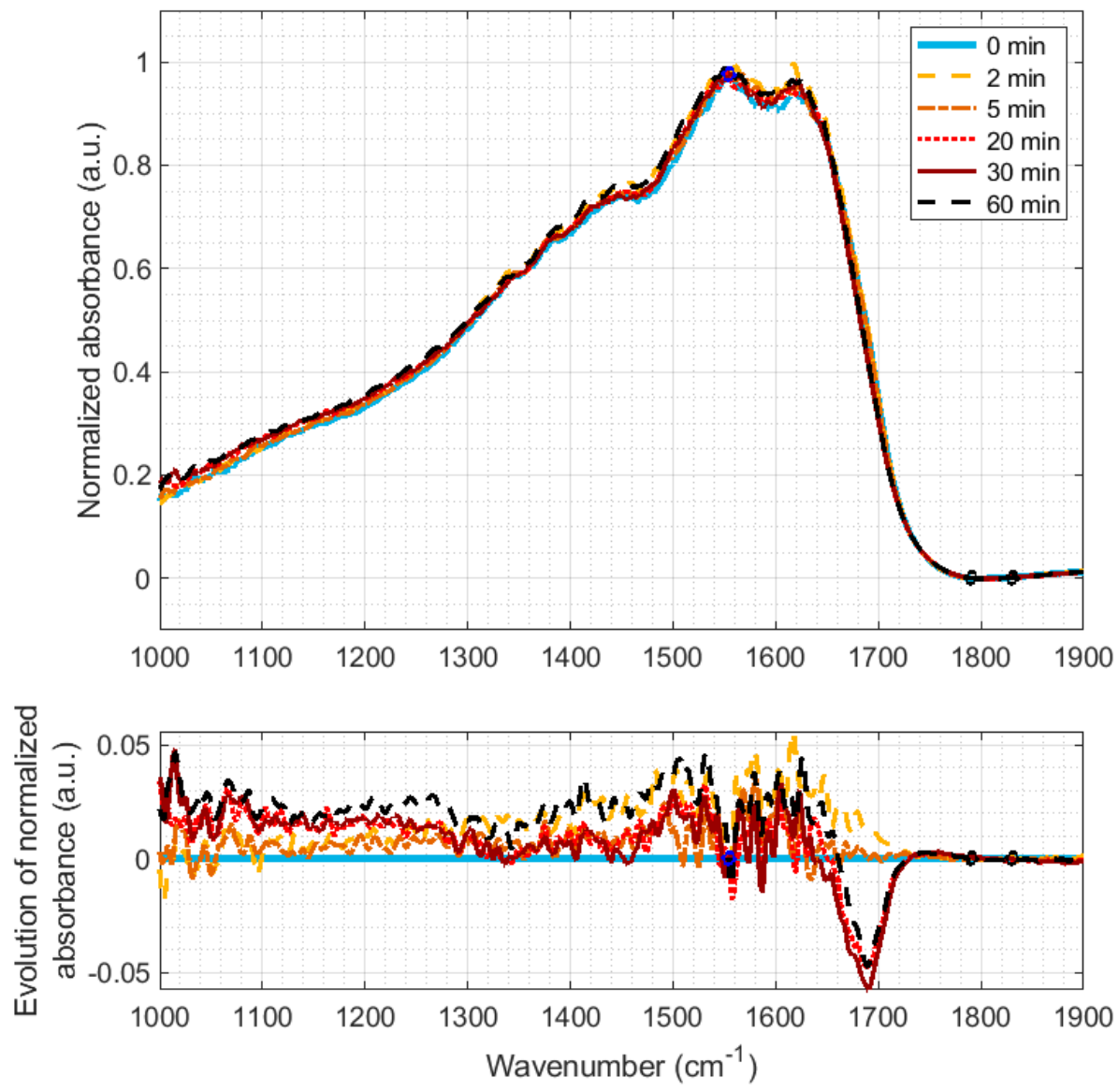

*Figure 16: Evolution of IR bands during heating of the sample $P_{gr}$ at 80-100°C at 4 mbar. Spectra of the three bands presented in the paper, 'CN' bands on Figure 7, 'CH' bands on Figure 8, and Figure 9.*

Only small evolutions are seen on IR spectra during sample heating: the shift of nitriles (discussed in section 3.3.2) and the loss of a band around 1690 $cm^{-1}$ due to VOCs desorption (discussed in section 3.3.3).

# 7. Acknowledgements

NC acknowledges the financial support of the European Research Council (ERC Starting Grant PRIMCHEM, Grant agreement no. 636829).

AC acknowledges ENS Paris-Saclay Doctoral Program.

# 8. Data availability

The spectra acquired for this study will be archived and accessible on the SPAN database (https://www.sshade.eu/db/span).

# 9. References

Abdu, Y.A., Hawthorne, F.C., Varela, M.E., 2018. Infrared Spectroscopy of Carbonaceous-chondrite Inclusions in the Kapoeta Meteorite: Discovery of Nanodiamonds with New Spectral Features and Astrophysical Implications. Astrophys. J. 856, L9. https://doi.org/10.3847/2041-8213/aab433

Ågren, K., Wahlund, J.-E., Garnier, P., Modolo, R., Cui, J., Galand, M., Müller-Wodarg, I., 2009. On the ionospheric structure of Titan. Planet. Space Sci. 57, 1821–1827. https://doi.org/10.1016/J.PSS.2009.04.012

Azzolina-Jury, F., Thibault-Starzyk, F., 2017. Mechanism of Low Pressure Plasma-Assisted CO2 Hydrogenation Over Ni-USY by Microsecond Time-resolved FTIR Spectroscopy. Top. Catal. 60, 1709–1721. https://doi.org/10.1007/s11244-017-0849-2

Bonnet, J.Y., Quirico, E., Buch, A., Thissen, R., Szopa, C., Carrasco, N., Cernogora, G., Fray, N., Cottin, H., Roy, L. Le, Montagnac, G., Dartois, E., Brunetto, R., Engrand, C., Duprat, J., 2015. Formation of analogs of cometary nitrogen-rich refractory organics from thermal degradation of tholin and HCN polymer. Icarus 250, 53–63. https://doi.org/10.1016/j.icarus.2014.11.006

Carrasco, E., Tanarro, I., Herrero, V.J., Cernicharo, J., 2013. Proton transfer chains in cold plasmas of H2with small amounts of N2. the prevalence of NH4+. Phys. Chem. Chem. Phys. 15, 1699–1706. https://doi.org/10.1039/c2cp43438e

Carrasco, N., Bonnet, J., Quirico, E., Thissen, R., Dutuit, O., Bagag, A., Lapre, O., Buch, A., Giulani, A., Adande, G., Ouni, F., Hadamcik, E., 2009. Chemical Characterization of Titan ' s Tholins : Solubility , Morphology and Molecular Structure Revisited † 11195–11203.

Carrasco, N., Jomard, F., Vigneron, J., Etcheberry, A., Cernogora, G., 2016. Laboratory analogues simulating Titan's atmospheric aerosols: Compared chemical compositions of grains and thin films. Planet. Space Sci. 128, 52–57. https://doi.org/10.1016/j.pss.2016.05.006

Carrasco, N., Tigrine, S., Gavilan, L., Nahon, L., Gudipati, M.S., 2018. The evolution of Titan's high-altitude aerosols under ultraviolet irradiation. Nat. Astron. 2, 1. https://doi.org/10.1038/s41550-018-0439-7

Coates, A.J., Crary, F.J., Lewis, G.R., Young, D.T., Jr, J.H.W., Jr, E.C.S., 2007. Discovery of heavy negative ions in Titan ' s ionosphere. Geophys. Res. Lett. 34, 22. https://doi.org/10.1029/2007GL030978

Courtin, R., 2005. Aerosols on the Giant Planets and Titan. Space Sci. Rev. 116, 185–199. https://doi.org/10.1007/s11214-005-1955-1

Cui, J., Yelle, R. V., Volk, K., 2008. Distribution and escape of molecular hydrogen in Titan's thermosphere and exosphere. J. Geophys. Res. E Planets 113, E10004. https://doi.org/10.1029/2007JE003032

D'Agostino, R., Favia, P., Fracassi, F., NATO Advanced Study Institute on Plasma Treatments and Deposition of Polymers (1996 : Acquafredda di Maratea, I., 1997. Plasma processing of polymers. Kluwer Academic Publishers.

Dartois, E., Muñoz Caro, G.M., Deboffle, D., Montagnac, G., d'Hendecourt, L., 2005. Ultraviolet photoproduction of ISM dust. Astron. Astrophys. 432, 895–908. https://doi.org/10.1051/0004-6361:20042094

Dionnet, Z., Aleon-Toppani, A., Baklouti, D., Borondics, F., Brisset, F., Djouadi, Z., Sandt, C., Brunetto, R., 2018. Organic and mineralogic heterogeneity of the Paris meteorite followed by FTIR hyperspectral imaging. Meteorit. Planet. Sci. 16, 1–16. https://doi.org/10.1111/maps.13178

Gautier, T., Carrasco, N., Mahjoub, A., Vinatier, S., Giuliani, A., Szopa, C., Anderson, C.M., Correia, J.J., Dumas, P., Cernogora, G., 2012. Mid- and far-infrared absorption spectroscopy of Titan's aerosols analogues. Icarus 221, 320–327. https://doi.org/10.1016/j.icarus.2012.07.025

Gronoff, G., Lilensten, J., Desorgher, L., Flückiger, E., 2009. Ionization processes in the atmosphere of Titan-I. Ionization in the whole atmosphere. Astron. Astrophys. 506, 955–964. https://doi.org/10.1051/0004-6361/200912371

Guerlet, S., Fouchet, T., Vinatier, S., Simon, A.A., Dartois, E., Spiga, A., 2015. Stratospheric benzene and hydrocarbon aerosols detected in Saturn's auroral regions. Astron. Astrophys. 580, A89. https://doi.org/10.1051/0004-6361/201424745

Hadamcik, E., Renard, J.B., Alcouffe, G., Cernogora, G., Levasseur-Regourd, A.C., Szopa, C., 2009. Laboratory light-scattering measurements with Titan's aerosols analogues produced by a dusty plasma. Planet. Space Sci. 57, 1631–1641. https://doi.org/10.1016/j.pss.2009.06.013

He, J., Buch, A., Carrasco, N., Szopa, C., 2015. Thermal degradation of organics for pyrolysis in space: Titan's atmospheric aerosol case study. Icarus 248, 205–212. https://doi.org/10.1016/j.icarus.2014.10.010

Hong, J., Truica-Marasescu, F., Martinu, L., Wertheimer, M.R., 2002. An Investigation of Plasma-Polymer Interactions by Mass Spectrometry. Plasmas Polym. 7, 245–260. https://doi.org/10.1023/A:1019938424698

Imanaka, H., Khare, B.N., Elsila, J.E., Bakes, E.L.O., McKay, C.P., Cruikshank, D.P., Sugita, S., Matsui, T., Zare, R.N., 2004. Laboratory experiments of Titan tholin formed in cold plasma at various pressures: Implications for nitrogen-containing polycyclic aromatic compounds in Titan haze. Icarus 168, 344–366. https://doi.org/10.1016/j.icarus.2003.12.014

Jacobson, M.C., Hansson, H.-C., Noone, K.J., Charlson, R.J., 2000. Organic atmospheric aerosols: Review and state of the science. Rev. Geophys. 38, 267–294. https://doi.org/10.1029/1998RG000045

Jia, Z., Rousseau, A., 2016. Sorbent track: Quantitative monitoring of adsorbed VOCs under in-situ plasma exposure. Sci. Rep. 6. https://doi.org/10.1038/srep31888

Kim, S.J., Jung, A., Sim, C.K., Courtin, R., Bellucci, A., Sicardy, B., Song, I.O., Minh, Y.C., 2011.

Retrieval and tentative indentification of the 3 μm spectral feature in Titans haze. Planet. Space Sci. 59, 699–704. https://doi.org/10.1016/j.pss.2011.02.002

Kuga, M., Marty, B., Marrocchi, Y., Tissandier, L., 2015. Synthesis of refractory organic matter in the ionized gas phase of the solar nebula. Proc. Natl. Acad. Sci. U. S. A. 112, 7129–34. https://doi.org/10.1073/pnas.1502796112

Laurent, B., Roskosz, M., Remusat, L., Leroux, H., Vezin, H., Depecker, C., 2014. Isotopic and structural signature of experimentally irradiated organic matter. Geochim. Cosmochim. Acta 142, 522–534. https://doi.org/10.1016/j.gca.2014.07.023

Lavvas, P., Galand, M., Yelle, R. V., Heays, A.N., Lewis, B.R., Lewis, G.R., Coates, A.J., 2011. Energy deposition and primary chemical products in Titan's upper atmosphere. Icarus 213, 233–251. https://doi.org/10.1016/j.icarus.2011.03.001

Lavvas, P, Sander, M., Kraft, M., Imanaka, H., 2011. Surface chemistry and particle shape: Processes for the evolution of aerosols in Titan's atmosphere. Astrophys. J. 728. https://doi.org/10.1088/0004-637X/728/2/80

Lavvas, P., Yelle, R. V., Koskinen, T., Bazin, A., Vuitton, V., Vigren, E., Galand, M., Wellbrock, A., Coates, A.J., Wahlund, J.E., Crary, F.J., Snowden, D., 2013. Aerosol growth in Titan's ionosphere. Proc. Natl. Acad. Sci. U. S. A. 110, 2729–2734. https://doi.org/10.1073/pnas.1217059110

Lebonnois, S. ébastie., Bakes, E.L.O., McKay, C.P., 2003. Atomic and molecular hydrogen budget in Titan's atmosphere. Icarus 161, 474–485. https://doi.org/10.1016/S0019-1035(02)00039-8

López-Puertas, M., Dinelli, B.M., Adriani, A., Funke, B., García-Comas, M., Moriconi, M.L., D'Aversa, E., Boersma, C., Allamandola, L.J., 2013. Large abundances of polycyclic aromatic hydrocarbons in Titan's upper atmosphere. Astrophys. J. 770, 132. https://doi.org/10.1088/0004-637X/770/2/132

Loureiro, J., Ricard, A., 1993. Electron and vibrational kinetics in an N2-H2 glow discharge with application to surface processes. J. Phys. D. Appl. Phys. 26, 163–176. https://doi.org/10.1088/0022-3727/26/2/001

Madeleine, J.-B., Forget, F., Millour, E., Montabone, L., Wolff, M.J., 2011. Revisiting the radiative impact of dust on Mars using the LMD Global Climate Model. J. Geophys. Res. 116, E11010. https://doi.org/10.1029/2011JE003855

Mohammad Gholipour, A., Rahemi, N., Allahyari, S., Ghareshabani, E., 2017. Hybrid Plasma-Catalytic Oxidation of VOCs with NiMn/Montmorillonite: Plasma and Catalyst Considerations. Top. Catal. 60, 934–943. https://doi.org/10.1007/s11244-017-0758-4

Morisson, M., Szopa, C., Carrasco, N., Buch, A., Gautier, T., 2016. Titan's organic aerosols: Molecular composition and structure of laboratory analogues inferred from pyrolysis gas chromatography mass spectrometry analysis. Icarus 277, 442–454. https://doi.org/10.1016/j.icarus.2016.05.038

Mutsukura, N., Akita, K.I., 1999. Infrared absorption spectroscopy measurements of amorphous CNxfilms prepared in CH4/N2r.f. discharge. Thin Solid Films 349, 115–119. https://doi.org/10.1016/S0040-6090(99)00237-0

Öberg, K.I., 2016. Photochemistry and Astrochemistry: Photochemical Pathways to Interstellar Complex Organic Molecules. Chem. Rev. https://doi.org/10.1021/acs.chemrev.5b00694

Pernot, P., Carrasco, N., Thissen, R., Schmitz-Afonso, I., 2010. Tholinomics—Chemical Analysis of Nitrogen-Rich Polymers. Anal. Chem. 82, 1371–1380. https://doi.org/10.1021/ac902458q

Pintassilgo, C.D., Guerra, V., 2017. Modelling of the temporal evolution of the gas temperature in N2discharges. Plasma Sources Sci. Technol. 26. https://doi.org/10.1088/1361-6595/aa5db2

Pöschl, U., 2005. Atmospheric aerosols: Composition, transformation, climate and health effects. Angew. Chemie - Int. Ed. https://doi.org/10.1002/anie.200501122

Quirico, E., Montagnac, G., Lees, V., McMillan, P.F., Szopa, C., Cernogora, G., Rouzaud, J.N., Simon, P., Bernard, J.M., Coll, P., Fray, N., Minard, R.D., Raulin, F., Reynard, B., Schmitt, B., 2008. New experimental constraints on the composition and structure of tholins. Icarus 198, 218–231. https://doi.org/10.1016/j.icarus.2008.07.012

Sciamma-O'Brien, E., Carrasco, N., Szopa, C., Buch, A., Cernogora, G., 2010. Titan's atmosphere: An optimal gas mixture for aerosol production? Icarus 209, 704–714. https://doi.org/10.1016/j.icarus.2010.04.009

Sekine, Y., Imanaka, H., Matsui, T., Khare, B.N., Bakes, E.L.O., McKay, C.P., Sugita, S., 2008a. The role of organic haze in Titan's atmospheric chemistry. I. Laboratory investigation on heterogeneous reaction of atomic hydrogen with Titan tholin. Icarus 194, 186–200. https://doi.org/10.1016/j.icarus.2007.08.031

Sekine, Y., Lebonnois, S., Imanaka, H., Matsui, T., Bakes, E.L.O., McKay, C.P., Khare, B.N., Sugita, S., 2008b. The role of organic haze in Titan's atmospheric chemistry. II. Effect of heterogeneous reaction to the hydrogen budget and chemical composition of the atmosphere. Icarus 194, 201–211. https://doi.org/10.1016/j.icarus.2007.08.030

Sharma, M.K., Saikia, B.K., Phukan, A., Ganguli, B., 2006. Plasma nitriding of austenitic stainless steel in N2 and N2-H2 dc pulsed discharge. Surf. Coatings Technol. 201, 2407–2413. https://doi.org/10.1016/j.surfcoat.2006.04.006

Šíra, M., Trunec, D., Sťahel, P., Buršíková, V., Navrátil, Z., Buršík, J., 2005. Surface modification of polyethylene and polypropylene in atmospheric pressure glow discharge. J. Phys. D. Appl. Phys. 38, 621–627. https://doi.org/10.1088/0022-3727/38/4/015

Socrates, 2004. Infrared and Raman characteristic group frequencies, Journal of Raman Spectroscopy. https://doi.org/10.1037/pspi0000025

Sode, M., Jacob, W., Schwarz-Selinger, T., Kersten, H., 2015. Measurement and modeling of neutral, radical, and ion densities in $H_2$-$N_2$-Ar plasmas. J. Appl. Phys. 117, 083303.

https://doi.org/10.1063/1.4913623

Szopa, C., Cernogora, G., Boufendi, L., Correia, J.J., Coll, P., 2006. PAMPRE: A dusty plasma experiment for Titan's tholins production and study. Planet. Space Sci. 54, 394–404. https://doi.org/10.1016/J.PSS.2005.12.012

Tanarro, I., Herrero, V.J., Islyaikin, A.M., Méndez, I., Tabare, F.L., Tafalla, D., 2007. Ion chemistry in cold plasmas of H2 with CH4 and N2. J. Phys. Chem. A 111, 9003–9012. https://doi.org/10.1021/jp073569w

Truica-Marasescu, F., Wertheimer, M.R., 2008. Nitrogen-rich plasma-polymer films for biomedical applications. Plasma Process. Polym. 5, 44–57. https://doi.org/10.1002/ppap.200700077

Vuitton, V., Yelle, R. V., Klippenstein, S.J., Hörst, S.M., Lavvas, P., 2019. Simulating the density of organic species in the atmosphere of Titan with a coupled ion-neutral photochemical model. Icarus 324, 120–197. https://doi.org/10.1016/j.icarus.2018.06.013

Waite, J.H., Young, D.T., Cravens, T.E., Coates, A.J., Crary, F.J., Magee, B., Westlake, J., 2007. Planetary science: The process of tholin formation in Titan's upper atmosphere. Science (80-. ). 316, 870–875. https://doi.org/10.1126/science.1139727

Wertheimer, M.R., Fozza, A.C., Holländer, A., 1999. Industrial processing of polymers by low-pressure plasmas: the role of VUV radiation. Nucl. Instruments Methods Phys. Res. Sect. B Beam Interact. with Mater. Atoms 151, 65–75. https://doi.org/10.1016/S0168-583X(99)00073-7

Westlake, J.H., Waite, J.H., Carrasco, N., Richard, M., Cravens, T., 2014. The role of ion-molecule reactions in the growth of heavy ions in Titan's ionosphere. J. Geophys. Res. Sp. Phys. 119, 5951–5963. https://doi.org/10.1002/2014JA020208

Zhang, X., West, R.A., Irwin, P.G.J., Nixon, C.A., Yung, Y.L., 2015. Aerosol influence on energy balance of the middle atmosphere of Jupiter. Nat. Commun. 6, 10231. https://doi.org/10.1038/ncomms10231